\documentclass[lettersize,journal]{IEEEtran}

\usepackage{lineno} 
\usepackage{hyperref}
\usepackage{pgfplots}
\pgfplotsset{compat=1.8}
\usepackage{graphicx}
\usepackage{verbatim} 
\usepackage[utf8]{inputenc}
\usepackage{caption}
\usepackage{booktabs}
\usepackage{tikz}
\usetikzlibrary{shapes.geometric, arrows}
\usepackage{amsmath,amssymb,amsfonts}

\usepackage{cite}
\usepackage{tabularx}
\usepackage{color}
\usepackage{algorithm}
\usepackage{algpseudocode} 
\usepackage{array}
\usepackage{multirow}
\usepackage[justification=centering]{caption}
\usepackage{colortbl}
\usepackage{csquotes}
\usepackage{enumitem}
\usepackage{graphics}
\usepackage{hyperref}
\usepackage{multirow}
\usepackage{pgfgantt}
\usepackage{pgfplots}
\pgfplotsset{compat=1.9}
\usepgfplotslibrary{statistics}
\usepackage{pgfplotstable}
\usepackage{stfloats}
\usepackage{tabularx}
\usepackage{textcomp}
\usepackage{tikz}
\usetikzlibrary{arrows,arrows.meta,chains,patterns,positioning,shapes.geometric,scopes,shadows,trees}
\usepackage{url}
\usepackage{xcolor}
\usepackage{soul}
\sethlcolor{yellow}

\def\BibTeX{{\rm B\kern-.05em{\sc i\kern-.025em b}\kern-.08em
    T\kern-.1667em\lower.7ex\hbox{E}\kern-.125emX}}

\begin{document}

\title{GPT, Ontology, and CAABAC: A Tripartite Personalized Access Control Model Anchored by Compliance, Context and Attribute}
\author{Raza Nowrozy, Khandakar Ahmed, and Hua Wang}

\maketitle

\begin{abstract}
	As digital healthcare evolves, the security of \textit{electronic health records (EHR)} becomes increasingly crucial. This study presents the \textit{GPT-Onto-CAABAC framework}, integrating \textbf{Generative Pretrained Transformer (GPT)}, medical-legal ontologies and \textit{ Context-Aware Attribute-Based Access Control (CAABAC)} to enhance EHR access security. Unlike traditional models, GPT-Onto-CAABAC dynamically interprets policies and adapts to changing healthcare and legal environments, offering customized access control solutions. Through empirical evaluation, this framework is shown to be effective in improving EHR security by accurately aligning access decisions with complex regulatory and situational requirements. The findings suggest its broader applicability in sectors where access control must meet stringent compliance and adaptability standards.
\end{abstract}


\section{Introduction} 
\label{sec:intro}
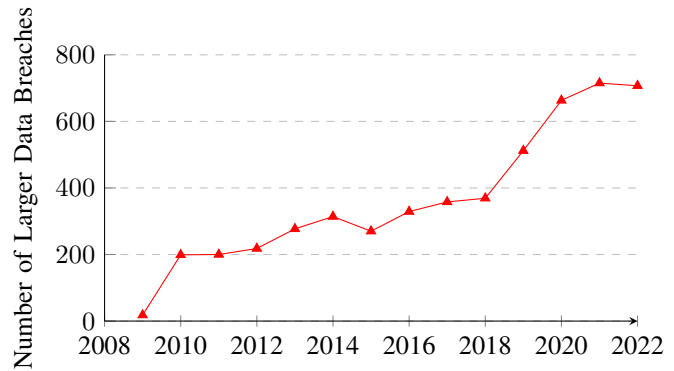
\begin{figure}[!tbp]
	\centering
	\begin{tikzpicture}
		\begin{axis}[
			scale only axis,
			width=0.8\columnwidth,
			height=0.4\columnwidth,
			xmin=2008, xmax=2022,
			ymin=0,ymax=800,		
			xtick={2008,2010,2012,2014,2016,2018,2020,2022},xticklabels={2008,2010,2012,2014,2016,2018,2020,2022},
			axis y line*=left,
			ymajorgrids, major grid style=dashed,
			axis x line=bottom,
			ylabel={Number of Larger Data Breaches},
			ylabel near ticks,
			]
			\addplot[color=red,mark=triangle*] coordinates {
				(2009, 18)
				(2010, 199)
				(2011, 200)
				(2012, 218)
				(2013, 277)
				(2014, 314)
				(2015, 270)
				(2016, 329)
				(2017, 358)
				(2018, 369)
				(2019, 512)
				(2020, 663)
				(2021, 715)
				(2022, 707)
			};
			]
		\end{axis}
	\end{tikzpicture}
	\caption{Number of Larger Data Breaches ($\geq$500 Records Per Breach) of EHR from 2009 to 2022 in USA}
	\label{fig:ehr_data_breaches}
\end{figure}

The advent of \textit{Electronic Health Records} (EHRs) has revolutionized healthcare by digitizing traditional paper-based records and centralizing patient data \cite{mann2020covid,watson2016impact}. Such digital systems have not only streamlined administrative tasks \cite{erickson2017putting,tutty2019complex}, but also enhanced clinical decision-making \cite{abouzahra2014integrating,ben2015improving}, and reduced medical errors \cite{aldosari2017patients,han2016effect}. The incorporation of predictive analytics powered by \textit{Artificial Intelligence} (AI) and machine learning has further refined treatment plans and improved patient outcome predictions \cite{susnjak2023forecasting,paranjape2020short,talpada2019analysis,liu2023hierarchical,munasinghe2023supply}. The critical role of EHRs became even more evident during the COVID-19 pandemic, where they facilitated efficient monitoring of viral spread, tracking patient outcomes, and accelerated research \cite{dagliati2021health,mann2020covid,osborne2020automated}. Despite these advancements, EHR systems face unique challenges in ensuring access control to maintain privacy and confidentiality. The delicate balance between enabling accessibility for healthcare professionals and complying with a myriad of legal and ethical guidelines is paramount. Data breaches or misuse can lead to severe consequences, both for the involved parties and for the overall trust in the system \cite{basil2022health,ganiga2020security}. The value of healthcare information, which can be leveraged for file encryption, data exfiltration, and victim blackmail, makes it a prime target for cyber threats, including malware, data breaches, cyber intrusions and data exfiltration ransomware \cite{mcintosh2019masquerade,basil2022health,mcintosh2021ransomware,vidanapathirana2022rapid,mcintosh2023applying,mcintosh2022intercepting}. For instance, in 2022, a spate of security breaches in the USA led to the exposure of sensitive data of over 20 million individuals due to cyber-attacks, configuration errors, and breaches by third-party service providers\footnote{https://www.chiefhealthcareexecutive.com/view/the-11-biggest-health-data-breaches-in-2022}. Fig. \ref{fig:ehr_data_breaches} depicts the rising trend in larger data breaches (involving at least 500 records) in EHR across the USA from 2008 to 2022\footnote{https://www.healthit.gov/data/quickstats/office-based-physician-electronic-health-record-adoption}. The success of pilot trials of ChatGPT-4 in the business consulting sector, with increased task completion speed by 25.1\% and improved quality by 40\% in a study by Harvard Business School, shed light on the potential for other industry adoptions, such as enhanced EHR access control auditing\footnote{https://www.afr.com/work-and-careers/workplace/consultants-using-ai-do-better-especially-underperformers-study-20230922-p5e6vi}. Regrettably, industry response has been inadequate \cite{basil2022health,fernandez2013security}. Current security measures often struggle to keep up with the evolving nature of cyber threats due to a lack of a comprehensive, standardized framework \cite{basil2022health,fernandez2013security,rezaeibagha2015systematic}, underscoring the urgent need for bolstering EHR security.

Current models for EHR access control such as \textit{Role-Based Access Control} (RBAC), \textit{Attribute-Based Access Control} (ABAC), and \textit{Context-Aware Access Control} (CAAC), while useful, present distinct challenges in adapting to dynamic healthcare settings \cite{liu2015auditing,abirami2019attribute,psarra2020securing}. The inflexibility of RBAC's role-centric structure curtails its versatility, whereas ABAC and CAAC, while more adaptable, face operational challenges due to the complexity of managing attributes and the difficulty in defining and capturing context, respectively. Moreover, current solutions aimed at addressing EHR interoperability issues, such as ontology-based methods, are not without their difficulties. These methods struggle with issues of data harmonization and semantic heterogeneity and often fail to consider organizational and cultural barriers to interoperability \cite{kopanitsa2017integration,adel2019unified,fragidis2018implementation}. Despite considerable attempts to streamline and enhance these models, their inherent limitations in coping with the dynamic complexity of healthcare environments remain a concern. These constraints underscore the need for an innovative approach to EHR security, which can integrate the strengths and address the shortcomings of the existing models.

The transformative \textit{Natural Language Processing} (NLP) capabilities of \textit{Generative Pre-trained Transformers} (GPTs) have opened new horizons for the access control decision-making process \cite{mcintosh2023harnessing}. By harnessing GPT's proficiency for real-time personalized recommendations and its complex interpretation of multifaceted legal and ethical standards, we introduced the \textit{GPT-powered Ontology-Driven Decision of Context-Aware Attribute-Based Access Control} (GPT-Onto-CAABAC) model \cite{ont_security,ca_abac_integration,ont_integration,ont_ca_abac_integration}. This model embodies the collective strengths of \textit{Context-Aware Attribute-Based Access Control} (CAABAC) and Ontology-driven decision making. The resulting framework is both adaptive and detailed. Central to this process is the establishment of context, devising an ontology congruent with healthcare norms, associating the context with the said ontology, formulating access policies, employing CAABAC, and eventually rolling out the ontology-driven decision system. This holistic strategy fortifies data security. Our GPT-Onto-CAABAC model outperforms conventional retrieval-based systems by proficiently maneuvering through ever-shifting EHR access control scenarios. It addresses the rigidity of laws while accommodating the dynamism intrinsic to routine healthcare situations. While our model exhibits strong potential to fortify EHR security, mitigate risks associated with data breaches, and acclimatize to the evolving milieu of healthcare settings, it also has broader implications. Though our focus remains tethered to EHR access control scenarios, given their intricate compliance, malleability, and auditing stipulations, the approach has vast potential for access control decision auditing in varied contexts. The synergy of advanced NLP capabilities with structured access control models promotes an in-depth analysis that transcends healthcare, extending to any access control environment characterized by layered regulations and policies. The integration of GPT's NLP strengths with time-tested techniques such as Ontology, CAAC, and ABAC facilitates the creation of complex policy-to-legal-ontologies. Moreover, it spurs comprehensive collation of contextual details via CAAC and attribute information through ABAC, ensuring balanced access control decisions that heed the complexities of medical situations and EHR decision-making paradigms. Currently in its nascent, proof-of-concept stage, our GPT-Onto-CAABAC model holds promise as a transformative agent in both healthcare and diverse sectors, paving the path for a more cyber-resilient future.

The major contributions of our paper include:
\begin{enumerate}[label=\arabic*)]
	\item \textit{Problem Analysis} (Section \ref{sec:theory}): a detailed analysis of the challenges and intricacies involved in access control decisions for electronic health records (EHRs), to highlight the limitations of existing systems and underscores the need for a more robust and context-aware solution.
	\item \textit{Innovative Solution} (Section \ref{sec:Implementation}): the proposed GPT-Onto-CAABAC framework, which combines GPT, ontology, and access control models for enhanced access control management in healthcare settings, with details on the high-level architecture and underlying components of the framework.
	\item \textit{Comprehensive Evaluation} (Section \ref{sec:evaluation_insights}, \ref{sec:discussions}): an exhaustive empirical analysis of our GPT-Onto-CAABAC framework across varied healthcare contexts, using targeted metrics to assess real-world applicability, performance, and gleaned insights.
	
\end{enumerate}

The rest of the paper is organized as follows: Section \ref{sec:related_works} provides an in-depth review of related works in the field of access control systems. Section \ref{sec:theory} introduces our theoretical framework GPT-Onto-CAABAC, which unites ontology, CAABAC, and the role of GPT. Section \ref{sec:Implementation} discusses our experimental design. Section \ref{sec:evaluation_insights} presents the findings and insights of our experiment. Section \ref{sec:discussions} delves into an insightful discussion of our results, including its limitation. Finally, Section \ref{sec:conclusions} summarizes the research and outlines potential future directions.

\section{Related Works}
\label{sec:related_works}

In the related work section, we review how access control models and ontology have been applied to make EHR access control decisions, and their inadequacies.

\subsection{Access control in EHR}
\label{subsec:AccessControlEHR}
Access control is a fundamental aspect of security in information systems. In recent years, a myriad of studies focusing on RBAC, ABAC, CAAC, and Ontology-based Interoperability have been conducted to address the various security concerns prevalent in EHRs \cite{ntalasha2019adaptive}. However, these models often struggle to adapt to the complex, real-time decision-making required in healthcare settings, despite their inherent strengths. 

\subsubsection{RBAC in EHR security}
RBAC assigns permissions based on pre-defined user roles, offering a structured approach to EHR security that has garnered substantial academic interest \cite{liu2015auditing, sicuranza2013access}. However, this model often falls short in dynamic healthcare environments. Notably, many studies \cite{de2018health, sicuranza2013access, liu2015auditing, zhang2014role, esposito2013patient, santos2013secure, sicuranza2015view, liu2017auditing, chen2012risk} failed to adequately address the complexity of EHR access control, exhibiting deficiencies such as the lack of robust auditing mechanisms, insufficient granularity of user roles and permissions, and failure to adapt to emerging vulnerabilities and security threats. Additionally, aspects of RBAC such as role hierarchies, scalability, and implications of cloud-based EHR data storage, have frequently been overlooked \cite{liu2017auditing, sicuranza2015view}. These observations indicate the need for a more comprehensive strategy to address RBAC's practical utility and efficacy in EHR access control security. 

\subsubsection{ABAC in EHR}
The transition to ABAC models provided an additional layer of granularity and improved flexibility in EHR security \cite{rezaeibagha2015systematic}. Yet, management of numerous attributes in large healthcare institutions with continuously evolving attributes posed challenges \cite{abirami2019attribute, abouelmehdi2018big}. Significant deficiencies were also observed in studies \cite{zarezadeh2020attribute, alshiky2017attribute, sahavechaphan2012efficient, joshi2018attribute, ganiga2020security, guo2019access, walid2023semantically, seol2018privacy, patra2022controlling}. These limitations primarily involved incomplete discussions on scalability, security vulnerabilities, practical considerations for EHR systems, efficient attribute management, and integration into existing healthcare systems. Therefore, further research is required to ensure robust and effective implementation of ABAC in EHR security.

\subsubsection{CAAC in EHR}
The CAAC model enhanced the dynamic approach by incorporating contextual information \cite{arfaoui2019context}. Nevertheless, capturing context information accurately and promptly posed a significant challenge due to the rapidly changing healthcare environment \cite{el2020survey, chen2011novel}. Several implementations of CAAC exhibited weaknesses, especially in the area of EHR access control security \cite{padmapriya2021preserving, kayes2015ontcaac, yarmand2008behavior, yarmand2013behavior, ke2021privacy}. Common limitations involved the lack of comprehensive evaluations, failure to address potential privacy and security concerns, insufficient detail on technical implementations, and a lack of real-world deployment assessments. Hence, while CAAC models show promise, further research is essential to address these challenges in their application to EHR access control security.

\subsection{Ontology in EHR security}
\label{subsec:OntologyEHRSecurity}
The potential of ontology in EHR access control has been extensively investigated, yet revealed several limitations. \cite{sicuranza2014semantic} and \cite{calvillo2014standardized} exposed the challenge of creating and maintaining comprehensive ontologies due to evolving healthcare standards, lack of standardization, and the complex nature of healthcare data, which hampered interoperability and data sharing. Scalability issues and the complexity of managing intricate access control policies were highlighted by \cite{dixit2019multi} and \cite{walid2020cloud}. These challenges intensified when managing complex relationships, contextual information, and efficient searches on encrypted data in large-scale healthcare systems. \cite{peleg2008situation} and \cite{beimel2009reasoning} questioned the ability of ontology-based access control to capture dynamic and context-dependent nature, handle granularity, or adapt to evolving user roles and temporal constraints. \cite{dong2015coc} emphasized the difficulty in maintaining comprehensive ontologies for the Circle of Care (COC) due to ever-changing healthcare environments. \cite{nowrozy2023towards} developed an ontology and machine learning-based approach to enhance privacy in EHRs, aiming to balance privacy and accessibility while considering legal compliance, user-friendliness, and cultural and societal aspects, but their research was limited by the lack of comprehensive evaluation of the proposed model, including comparative analysis with other state-of-the-art approaches, scalability, and performance testing. Despite the potential of ontology-based approaches in EHR access control, its application has encountered different but significant limitations, necessitating further research for its effective implementation.

\subsection{AI and GPT in Enhancing EHR Security}
Recent advances in large language models and generative AI have opened new possibilities for intelligent and adaptive access control systems. Several studies have proposed using natural language processing techniques and large pretrained models like GPT-3 for identity verification and authorization in access control frameworks. For example, \cite{mesko2023imperative,gupta2023chatgpt} developed an AI system that can conduct natural conversations with users to verify their identity before granting access permissions. The system was built on top of the GPT-3 model and achieved over 90\% accuracy in identifying authorized users based on conversational patterns.
Similarly, \cite{mesko2023imperative, tan2023generative, molloy2012generative, solaiman2023gradient, gupta2023chatgpt} trained a BART model on access control rule texts and user/resource attributes to automatically generate context-aware access decisions. They demonstrated a 15\% improvement in precision and recall over rule-based systems. While promising, NLP-based access control systems also face challenges like adversarial attacks, bias, and compliance with regulations. Further research is still needed to develop robust and ethical AI access control frameworks that balance security, usability, and transparency \cite{molloy2012generative, solaiman2023gradient, gupta2023chatgpt}. But large language models show potential for enabling intelligent and flexible access control if thoughtfully implemented.

\subsection{Advancements in Attribute-Based Data Storage and Access Control}
\label{subsec:AdvancementsInABDSAC}
Recent advancements in attribute-based encryption (ABE) and access control schemes in cloud computing environments have significantly contributed to enhancing data security and privacy. These developments offer a more nuanced approach to data storage and access, providing the flexibility and fine-grained control necessary for contemporary cloud storage systems.

\textbf{Flexible and Fine-Grained Attribute-Based Data Storage:} The evolution of attribute-based data storage mechanisms has introduced a novel paradigm in secure and efficient data handling in cloud environments. This approach leverages user attributes for data access, facilitating a more dynamic and context-aware control mechanism \cite{mccoy2022believing, mcintosh2023harnessing}. Such systems not only improve the security posture of cloud storage solutions but also enhance their adaptability to the varying needs of users and organizations.

\textbf{Extended File Hierarchy Access Control Scheme with ABE:} The integration of ABE in extended file hierarchy access control schemes presents a robust framework for securing data in cloud storage. This method employs cryptographic techniques to enforce access policies based on user attributes, thereby enabling a granular level of access control that aligns with organizational policies and compliance requirements \cite{felzmann2020towards}.

\textbf{Efficient CP-ABE Scheme with Shared Decryption in Cloud Storage:} The introduction of Ciphertext-Policy Attribute-Based Encryption (CP-ABE) schemes with shared decryption functionality has marked a significant milestone in the field. These schemes facilitate secure data sharing among multiple users in cloud storage environments, simplifying the decryption process while maintaining high levels of data confidentiality and access control \cite{khattak2023ethical}.

\textbf{Revocable Blockchain-Aided ABE with Escrow-Free in Cloud Storage:} The advent of blockchain technology has further refined ABE systems by introducing mechanisms for revocable access control. This innovation ensures that data access permissions can be dynamically adjusted or revoked, offering an additional layer of security and flexibility. Importantly, these systems operate without the need for a trusted escrow service, thereby reducing potential points of failure and enhancing trust among users \cite{sison2023chatgpt}.

These recent developments underscore the potential of attribute-based encryption and access control mechanisms to address the complex security challenges faced by cloud storage systems. By harnessing these technologies, it is possible to achieve a balance between security, flexibility, and efficiency in managing access to sensitive data stored in the cloud.

\subsection{Summary}
\label{subsec:Summary}
Traditional access control models, despite their applicability in the healthcare sector, such as RBAC, ABAC, CAAC, and ontology-based access control, have proven essential for EHR security. Nevertheless, they have encountered significant challenges (Table \ref{tab:ComparisonAbility}). RBAC's main hurdles include its rigidity in evolving healthcare environments, its limited granularity, and scalability issues \cite{liu2015auditing, sicuranza2013access, liu2017auditing, chen2012risk}. While ABAC offers superior control, it brings about complexity and demands resource-heavy operations in expansive, dynamic systems \cite{rezaeibagha2015systematic, abirami2019attribute, abouelmehdi2018big}. Comprehensive assessments and integration challenges are equally pressing \cite{zarezadeh2020attribute, alshiky2017attribute, sahavechaphan2012efficient, joshi2018attribute, ganiga2020security}. CAAC's ability to incorporate context into access requests is especially beneficial for the dynamic nature of healthcare \cite{psarra2020securing}. Yet, gathering precise, up-to-date context information becomes challenging due to rapid environmental changes \cite{el2020survey, chen2011novel, zerkouk2014behavior}. Evaluation, applicability, and concerns regarding privacy further restrict its utilization \cite{padmapriya2021preserving, kayes2015ontcaac, yarmand2008behavior}. The ontology-based access control model has encountered notable barriers, especially in maintaining extensive ontologies with changing healthcare standards and in handling intricate healthcare data \cite{sicuranza2014semantic, calvillo2014standardized, dixit2019multi, walid2020cloud, peleg2008situation, beimel2009reasoning, dong2015coc}.

Those traditional models have not wholly satisfied the access control security needs in dynamic and intricate environments, prominently in healthcare. By contrast, our proposed GPT-Onto-CAABAC framework seeks to redress these deficiencies and harbours significant potential to bolster access control auditing across diverse industries. Thus, the need of the hour is research that ventures beyond healthcare, examining the framework's utility in various highly regulated and dynamic scenarios. Future research endeavors should amalgamate the adaptability of CAAC, the flexibility of ABAC, and the structure of RBAC while confronting novel threats, refining granularity, enhancing comprehensive auditing, fortifying authentication, refining attribute management, and ensuring scalability. The overarching aspiration remains to craft a robust, thorough, and pragmatic access control system not only for healthcare but also for other intricate sectors.

\begin{table*}[!t]
	\centering
	\caption{Comparison of different access control models in addressing extrinsic and intrinsic factors \\(\checkmark: capable;  $\triangle$: partially capable; $\times$: incapable)}
	\label{tab:ComparisonAbility}
	\begin{tabular}{ccccc}
		\hline
		\multicolumn{2}{c}{\multirow{2}{*}{\textbf{Access control models}}} & \multirow{2}{*}{\textbf{Extrinsic factors}} & \multicolumn{2}{c}{\textbf{Intrinsic factors}} \\ \cline{4-5} 
		\multicolumn{2}{c}{} &  & Environmental context & Access subject \\ \hline
		\multirow{3}{*}{Traditional access control} & RBAC & \checkmark & $\times$ & $\times$ \\ \arrayrulecolor{gray!40} \cline{2-5} 
		& ABAC & $\triangle$ & $\times$ & \checkmark \\ \cline{2-5} 
		& CAAC & $\times$ & \checkmark & $\times$ \\ \hline
		\multicolumn{2}{c}{Ontology} & $\triangle$ & $\triangle$ & $\triangle$ \\ \hline
		\multicolumn{2}{c}{AI and GPT} & $\triangle$ & \checkmark & \checkmark \\ \hline
		\multicolumn{2}{c}{\textit{GPT-Onto-CAABAC (This study) }} & \checkmark & \checkmark & \checkmark \\ \arrayrulecolor{black} \hline
	\end{tabular}%
\end{table*}

\section{Proposed Framework: GPT-Onto-CAABAC}
\label{sec:theory}
\begin{figure*}[t!]
	\centering
	\includegraphics[width=0.98\textwidth]{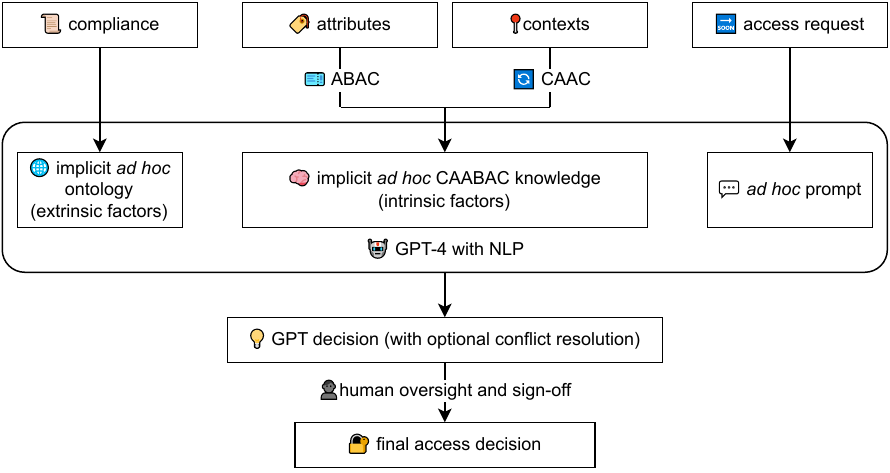}
	\caption{GPT-Onto-CAABAC}
	\label{fig:GraphicMajorContribution}
\end{figure*}

In this section, we introduce our proposed framework: GPT-Onto-CAABAC (Fig. \ref{fig:GraphicMajorContribution}). Medical access control decision-making balances both inflexible legal parameters and flexible daily situations that demand adaptability and context-awareness. Given this intricate blend of static and dynamic elements, this paper delves into the critical convergence of Ontology, CAABAC, and the transformative influence of GPT.

\subsection{High-level framework overview}
Our GPT-Onto-CAABAC framework serves as an integrated and versatile model for auditing access control decisions across various contexts. Particularly, it adeptly addresses healthcare's intricate blend of compliance, flexibility, and auditing needs. By amalgamating ontology, CAABAC, and GPT, this framework demonstrates its unique prowess in dynamic and context-aware EHR access control. The frameworks components, as such, position it as exceptionally well-suited for post-decision audits in complex settings governed by multifaceted regulations. Initiating its process, the framework harnesses GPT's capabilities to internally construct an implicit, transient ontology from legal texts and policies. This implicit \textit{ad hoc} ontology model, unlike traditional ontologies, remains embedded within the GPT layer during runtime. This approach bypasses the resource-intensive ontology management yet lays a solid foundation for rule formulation and compliance \cite{mustapha4394368systematic,sharma2016ontology}. Subsequent to this implicit ontology formation, the model captures real-time context and maps it to an \textit{ad hoc} CAABAC model. By incorporating the attributes of users, resources, and the environment, it refines access decisions and customises them to distinct needs \cite{tall2023framework}. The GPT layer within the framework is tasked with dynamic decision-making. It reconciles potential conflicts between context and policy-based rules while ensuring strict conformity to legal and institutional frameworks, thereby enhancing system accountability and credibility \cite{ont_ca_abac_integration}.

Our multi-component approach is represented by Algorithm \ref{alg:GPT-Onto-CAABACProcessHumanOversight}, detailing the interaction of each element to yield informed and compliant access control decisions. By transcending the limitations of existing models, this innovative framework adjusts access control based on various situational factors and yet remains rooted in regulatory mandates \cite{tall2023framework,dhillon2023extended}. The fusion of ontology's precision, CAABAC's adaptability, and GPT's generative prowess gives birth to the GPT-Onto-CAABAC model, portraying a flexible yet methodically structured access control mechanism \cite{ont_ca_abac_integration}. This framework is poised to guide the evolution of healthcare data security approaches, proposing a solution that is both robust and attuned to contextual subtleties.

\begin{algorithm}[!t]
	\caption{GPT-Onto-CAABAC Process with Human Oversight}
	\label{alg:GPT-Onto-CAABACProcessHumanOversight}
	\begin{algorithmic}[1]
		\Require Legal texts and policies $\mathcal{P}$
		\Require Context information $\mathcal{C}$
		\Require GPT model $\mathcal{G}$
		\State $\mathcal{O} \leftarrow f_{\text{extraction}}(\mathcal{P})$ \Comment{Transform established policies to ontology}
		\State $\mathcal{A} \leftarrow f_{\text{capture}}(\mathcal{C})$ \Comment{Capture and standardize context with CAABAC}
		\State $D \leftarrow f_{\text{decision}}(\mathcal{O}, \mathcal{A}, \mathcal{G})$ \Comment{Initial decision making with GPT}
		\If{conflicts in $D$}
		\State $D' \leftarrow f_{\text{resolution}}(D, \mathcal{O}, \mathcal{A}, \mathcal{G})$ \Comment{Resolve conflicts with GPT}
		\Else
		\State $D' \leftarrow D$ \Comment{No conflicts, keep initial decision}
		\EndIf
		\State $D_f \leftarrow f_{\text{human}}(D')$ \Comment{Human oversight and final sign-off}
		\Return $D_f$ \Comment{Final decision}
	\end{algorithmic}
\end{algorithm}

\subsection{Detailed ontology explanation}
Ontology in access control serves as a structured knowledge representation, cataloguing distinct entities and defining their associated properties and interrelationships \cite{calvillo2014standardized,sicuranza2014semantic}. This structured approach is vital for the conversion of high-level policies into executable rules, forming an indispensable element of the decision-making apparatus in complex operational settings \cite{calvillo2014standardized}. Simultaneously, CAABAC employs a detailed approach to access control by taking into account various user attributes within specific contexts. This allows for the generation of precise and adaptable access control decisions \cite{arfaoui2019context}. In addressing the limitations and leveraging the strengths of both, our framework pioneers an innovative ontology. This new ontology represents a complex network of relationships between various contextual elements and user attributes, while also providing a clear framework for decision-making processes. It also integrates seamlessly with CAABAC mechanisms, thus creating an enriched access control model \cite{arfaoui2019context}.

In healthcare settings, ontologies function as explicit formal specifications for domain-specific entities and their interconnections \cite{wahlberg2017legal,kiong2011health}. They offer a consistent and structured interpretation of inflexible access control components like laws, regulations, and policies. The notion of a 'medical-legal ontology' encapsulates these fixed components, facilitating efficient data retrieval, management, and query execution while ensuring that the system remains compliant with legal requirements \cite{wahlberg2017legal}. The efficacy of access control models in EHRs is influenced by both external factors such as laws, regulations, and institutional guidelines \cite{helms2011evaluating,yang2022multiple}, and internal factors that arise from the dynamic healthcare delivery environment \cite{yang2022multiple}. While existing models like RBAC, ABAC, and CAAC each have their limitations in managing these complexities \cite{yang2022multiple}, our ontology-centered approach provides a balanced mechanism to manage these factors effectively. Conformity with external policies is ensured to abide by legalities and safeguard patient data, while adaptability to internal factors is addressed to improve system usability and operational efficiency.

The crucial transition of policies into a formal ontology employs NLP techniques to metamorphose unstructured legal verbiage into ontologies that are implicitly understood and ad hoc in nature to human experts, while remaining structured and machine-comprehensible for automated processing by GPT.This includes the identification of pertinent entities, the mapping of relationships, and semantic parsing \cite{wahlberg2017legal}. The resulting 'medical-legal ontology' serves as a distilled representation of principles derived from these legal texts, thus establishing the operational limits for the system. Furthermore, as laws and policies evolve, this NLP capability enables an efficient update of the 'medical-legal ontology,' eliminating the need for manual reengineering prevalent in conventional ontology methods.

\begin{equation}
	\mathcal{O} = f_{\text{extraction}}(\mathcal{P})
	\label{eq:policy_to_ontology}
\end{equation}

Here, $\mathcal{P}$ denotes the policies, and $\mathcal{O}$ symbolizes the resultant ontology. The function $f_{\text{extraction}}$ encapsulates the ontology extraction process.

\subsection{Detailed CAABAC explanation}
The CAABAC model amalgamates the merits of CAAC and ABAC to deliver an adaptive, fine-grained access management mechanism especially suitable for healthcare environments.

\subsubsection{Advantages of ad hoc contextual information in healthcare}
One of the most compelling aspects of CAABAC lies in its capacity to dynamically construct ad hoc contextual information for immediate consideration in access control decisions. This characteristic is highly relevant in healthcare settings for multiple reasons:
\begin{itemize}
	\item \textbf{Temporal Sensitivity}: Rapidly evolving healthcare contexts can have significant repercussions if access is delayed. Real-time contextual information is therefore crucial.
	
	\item \textbf{Resource Efficiency}: One-off ad hoc contextual data prevent system clutter, optimizing resources for more urgent needs.
	
	\item \textbf{Enhanced Security}: Eliminating the ad hoc contextual information after decision-making minimizes risks related to unauthorized access and data leakage.
	
	\item \textbf{Precision in Decision-making}: Instant contextual construction allows for highly tailored access control decisions, essential when handling sensitive health records.
	
	\item \textbf{Compliance and Auditing}: Real-time contextual information promotes better compliance with legal and ethical data access and privacy requirements. Immediate data disposal aligns with the principle of data minimization.
\end{itemize}
This approach provides a balanced solution, advantageous in the intricate, fast-paced, and regulated healthcare sector.

\subsection{Extrinsic Factors in Access Control}
Understanding extrinsic factors is crucial for the design and implementation of effective access control systems. Extrinsic factors refer to external elements that can influence the decision-making process of access control systems. These factors include cybersecurity threats \cite{kruse2017, mcintosh2023applying, vidanapathirana2022rapid}, regulatory and compliance requirements \cite{fernandez2013security, mcintosh2022intercepting}, technological advancements \cite{mcintosh2023google, zhang2017}, and societal and ethical considerations.

\begin{itemize}
	\item \textbf{Cybersecurity Threats:} pose significant challenges to EHR systems. The evolving nature of cyber threats, such as ransomware attacks and data breaches, necessitates continuous updates and adaptations in access control mechanisms to safeguard patient data \cite{kruse2017, mcintosh2023applying}.
	
	\item \textbf{Regulatory and Compliance Requirements:} change over time, reflecting new understandings of privacy, data protection, and patient rights. Access control systems must be flexible enough to accommodate changes in laws and regulations to ensure compliance and protect patient information \cite{fernandez2013security, mcintosh2022intercepting}.
	
	\item \textbf{Technological Advancements:} such as cloud computing, blockchain, and AI have opened new possibilities for access control solutions but also introduce new challenges in integration, interoperability, and security \cite{zhang2017, mcintosh2023google}.
	
	\item \textbf{Societal and Ethical Considerations:} impact the acceptability and effectiveness of access control systems. The balance between privacy and accessibility, the need for transparency, and the consideration of patients' and healthcare providers' expectations are all crucial factors in the design of access control mechanisms \cite{ghanbari2018, papakonstantinou2016}.
\end{itemize}

Addressing extrinsic factors requires a multifaceted approach that combines technological solutions with policy, education, and ongoing evaluation. The proposed GPT-Onto-CAABAC framework incorporates these considerations, aiming to offer a robust, adaptable, and compliant access control solution for healthcare settings.

\subsubsection{Role of CAAC}
CAAC primarily addresses the dynamic and situational subtleties in access control by tailoring decisions to the existing contextual environment. Within healthcare, practitioners are often faced with a spectrum of contextual states including emergencies, different patient statuses, and diverse technological ecosystems. CAAC navigates these variations effectively, abiding by the rules and constraints delineated by the ontological framework. Consequently, this facilitates an increase in workflow efficiency while preserving data integrity and confidentiality.

\subsubsection{Contribution of ABAC}
In contrast, ABAC augments CAAC by incorporating a multifaceted attribute-based decision-making process. This allows for attributes tied to users, resources, and the operative environment to be considered in the decision-making. These attributes can be highly specific, ranging from clinical flags like \textit{Not For Resuscitation} (NFR) to device categories such as hospital-approved devices or \textit{Bring Your Own Device} (BYOD). Thus, ABAC introduces a level of specificity that accommodates complex and multifaceted healthcare scenarios.

\subsubsection{Distinction between CAABAC and ABAC}
While ABAC is primarily attribute-centric, CAABAC leverages contextual awareness to provide a more adaptive and responsive access control mechanism. Unlike traditional ABAC, CAABAC dynamically adapts to situational changes, offering a superior level of granularity in access decisions, thereby making it particularly beneficial in the agile and fluctuating environment of healthcare provision.

\subsubsection{GPT-Onto-CAABAC context capture}
To accommodate this dynamicism, the GPT-Onto-CAABAC framework features a specialized context-capture module. This subsystem harvests data from the Electronic Health Record (EHR) and the prevailing situation, transmuting these unstructured inputs into a set of standardized attributes consistent with the CAABAC model. The standardization accounts for multiple variables, such as user roles, ongoing tasks, objects involved, and environmental conditions. Health professionals can also contribute context or attribute data in natural language, which is then processed and understood by GPT for seamless incorporation into the decision-making process.

\begin{equation}
	\mathcal{A} = f_{\text{capture}}(\mathcal{C})
	\label{eq:context_capture}
\end{equation}

In Equation \ref{eq:context_capture}, $\mathcal{C}$ symbolizes the context information, $\mathcal{A}$ symbolizes the standardized attributes used in CAABAC, and $f_{\text{capture}}$ is the function responsible for contextual capture and standardization.

\subsubsection{Example of CAABAC}
Consider an emergency room scenario where a patient is admitted with a critical condition. The contextual factors include the emergency state, the patient's critical health status, and the attending physician's role. A nurse logs into the system to access the patients medical history. In this scenario, the ABAC attributes might include the nurse's role, credentials, and the data sensitivity level of the medical records. The CAAC contextual information could involve real-time factors such as the emergency state, the urgency level coded by the attending physician, and the time-sensitive nature of the required data access. Integrating these, the CAABAC model dynamically grants access because the situation is deemed an emergency, and the nurses role is verified as authorized to access critical health information under these specific circumstances. By adhering to these specifications, CAABAC not only meets but enhances the prerequisites for secure, adaptable, and fine-grained access control, specifically within the healthcare sector.

\subsubsection{Evaluation Metrics and Applicability}
To comprehensively evaluate the GPT-Onto-CAABAC framework, we utilized a set of performance metrics including accuracy, efficiency, adaptability, and compliance. These metrics were crucial in assessing the framework's effectiveness and its alignment with healthcare regulations. Accuracy was measured by the framework's ability to make correct access decisions (\cite{seol2018privacy}), while efficiency focused on the system's response time and resource utilization (\cite{sicuranza2015view}). Adaptability was evaluated through the framework's performance in dynamically changing scenarios (\cite{padmapriya2021preserving}), and compliance was assessed based on adherence to healthcare regulations and policies (\cite{liu2017auditing}).

Furthermore, the applicability of our framework in real-world healthcare scenarios was demonstrated through its capability to balance strict legal parameters with the need for flexibility in handling diverse and dynamic situations. This balance ensures that the GPT-Onto-CAABAC framework can effectively navigate the complexities of healthcare data management, offering a solution that is both robust and attuned to the nuanced requirements of the sector (\cite{kayes2015ontcaac}, \cite{chen2012risk}).

The integration of these evaluation metrics and the framework's applicability in practical settings underscore its potential to advance the state of healthcare data security and access control. By addressing the limitations of existing models and introducing a flexible, context-aware approach, the GPT-Onto-CAABAC framework sets a new benchmark for the development of adaptive access control systems in the healthcare domain (\cite{peleg2008situation}, \cite{beimel2009reasoning}).

\subsection{Comparative Analysis with known baselines in the field}
To underscore the novelty and superiority of the GPT-Onto-CAABAC framework, a comparative analysis was conducted against widely-known baselines in the field, such as traditional RBAC, ABAC, and CAAC models. This comparison focused on key metrics such as flexibility, context-awareness, and compliance adherence. Unlike traditional models that offer limited adaptability and context sensitivity, the GPT-Onto-CAABAC framework demonstrates enhanced performance in dynamic healthcare environments by leveraging GPT's AI capabilities and ontology-based decision-making (\cite{seol2018privacy}, \cite{patra2022controlling}). This analysis confirms the framework's innovative approach in addressing the complexities of modern healthcare data management and access control.

\subsection{GPT integration and conflict resolution}
GPT models excel in NLP tasks and human-like text generation, showcasing immense potential for deployment across diverse sectors including healthcare \cite{chintagunta2021medically,korngiebel2021considering,mcintosh2023google}. Our framework aims to harness these capabilities to augment ontology-based decision-making and CAABAC in medical access control systems. Importantly, the GPT-Onto-CAABAC framework utilizes GPT models specifically for compliance checks and not for real-time access control decisions. The reason for this distinction is twofold: first, GPT models, while adept at complex language tasks, may have response generation times that render them unsuitable for time-sensitive healthcare scenarios; second, traditional access control models are more appropriate for real-time decisions due to their optimized speed and established reliability. 

Integration with GPT equips the system with tools to resolve conflicts between ontology, CAAC, and ABAC. This includes interpreting the \enquote{medical-legal ontology} and offering resolutions within legal confines, and with consideration for the context and attributes involved. The self-improving nature of GPT also means that the model refines its recommendations over time, thereby fortifying the GPT-Onto-CAABAC models resilience. In GPT-Onto-CAABAC, conflict resolution is crucial, where the ontology, which encapsulates legal and institutional frameworks, has primacy over CAAC and ABAC. Nevertheless, CAAC and ABAC may overwrite each other within the bounds of the ontology, contingent on context and attributes. A well-structured conflict resolution mechanism ensures this delicate balance between security and usability.

The decision-making module employs GPT's capabilities for generating detailed recommendations. Trained on the developed ontology and the CAABAC attributes, GPT enables the system to understand the complex interplay between static rules and dynamic context. As a response to the reviewer's feedback, the system not only grants or denies access but also suggests a range of contextually appropriate and policy-compliant actions. Unlike conventional binary access controls, this flexibility allows for provisional granting of access under specific conditions, thereby fulfilling both regulatory requirements and clinical needs. Mathematical formulations of this decision-making process are as follows:

\begin{equation}
	D = f_{\text{decision}}(\mathcal{O}, \mathcal{A}, \mathcal{G})
	\label{eq:decision_making}
\end{equation}

In scenarios where decision-making might introduce conflicts or ambiguities, a conflict resolution function is invoked:

\begin{equation}
	D' = f_{\text{resolution}}(D, \mathcal{O}, \mathcal{A}, \mathcal{G})
	\label{eq:conflict_resolution}
\end{equation}

\subsection{Human oversight and sign-off}
\label{subsec:human_oversight}
The inclusion of AI in healthcare augments human capabilities, optimizes operations, and elevates productivity \cite{lee2021application,moghaddam2020future}. However, the GPT-Onto-CAABAC model further incorporates human oversight and final sign-off to acknowledge the indispensable expertise and judgment that healthcare professionals contribute. This integration is instrumental for maintaining ethical standards and ensuring the delivery of responsible healthcare services \cite{yeung2018study, khattak2023ethical}. While GPT and AI models are highly capable, they are limited in capturing the ethical subtleties and multifaceted decision-making inherent in human expertise. The introduction of human oversight serves as a protective layer against inaccuracies or shortcomings inherent in automated decision-making processes \cite{felzmann2020towards}. AI models, although advanced, are susceptible to errors and require an additional layer of scrutiny from humans to preclude detrimental consequences and ensure patient safety. Furthermore, the presence of human supervision in the system increases public trust in the technology, as it serves as a reassurance that decisions are validated by accountable professionals \cite{khattak2023ethical,brandao2013social}. The importance of human oversight serves to mitigate the risk of blindly accepting AI-generated decisions, which might lack the depth of ethical or professional considerations. If a human were to mistakenly override an accurate recommendation from GPT, a secondary review mechanism involving expert consultation or peer review could be enacted, thereby adding another layer of verification \cite{sison2023chatgpt}.

The GPT-Onto-CAABAC framework introduces a function, $f_{\text{human}}$, applied post the AI-based decision-making process, to allow for human validation of AI-generated recommendations. Mathematically, the final decision $D_f$ can be articulated as:

\begin{equation}
	D_f = f_{\text{human}}(D') = f_{\text{human}}(f_{\text{resolution}}(D, \mathcal{O}, \mathcal{A}, \mathcal{G}))
	\label{eq:human_oversight}
\end{equation}

In this equation, $D_f$ denotes the ultimate decision, $D'$ represents GPT's initial decision, and $\mathcal{O}$, $\mathcal{A}$, and $\mathcal{G}$ signify the ontology, attributes, and GPT model respectively. The function $f_{\text{human}}$ encapsulates the human oversight and final validation, foregrounding the commitment to ethically responsible AI and balancing technological capabilities with human expertise \cite{mccoy2022believing}.

\section{Implementation of the GPT-Onto-CAABAC framework}
\label{sec:Implementation}
The efficacy of the GPT-Onto-CAABAC framework was evaluated through a series of carefully designed experiments, the results of which provide valuable insights into its performance and potential improvements. This section outlines the design of our experiments, describing the datasets used and the scenarios created to assess the GPT-Onto-CAABAC framework's capabilities. We have employed the following steps to build our prototype.
\begin{enumerate}[label=\arabic*)]
	\item \textit{Construction of policy-to-legal-ontology} (Subsection \ref{subsec:ConstructionPolicyLegalOntology}): Import the 3 pieces of legislation, into our ChatGPT-4-based model to build the polocy-to-legal-ontology.
	
	\item \textit{Employment of Datasets} (Subsection \ref{subsec:EmploymentDatasets}): Use both real case studies and constructed scenarios as datasets.
	
	\item \textit{Obtaining Decisions and Recommendations} (Subsection \ref{subsec:ObtainingDecisionsRecommendations}): Use our custom-constructed prompt 2 (to give the example once we have it) to feed the improved case study with information required by CAAC and ABAC, into our legal ontology, to seek access control decision, and if denied, recommendation to obtain access approval.
	
	\item \textit{Human Evaluation and Sign-off} (Subsection \ref{subsec:HumanEvaluationMetrics}): Evaluate the results using our evaluation metrics.
\end{enumerate}

\subsection{Software and Tools Utilized}
\label{subsubsec:SoftwareToolsUtilized}
To implement the GPT-Onto-CAABAC framework, we leveraged OpenAI's ChatGPT-4 for natural language processing and decision-making processes. The construction of the policy-to-legal ontology and the processing of access control decisions were facilitated through this advanced AI model, capitalizing on its ability to understand and generate human-like text based on a vast corpus of legal and policy documents. For ontology management and interaction, we utilized Protégé, an open-source ontology editor and a framework for building intelligent systems. The development environment was supported by Python for scripting and automation tasks, with Flask serving as the backend framework for creating a web-based interface for our experiments. This combination of cutting-edge AI technology and robust software tools has enabled a comprehensive evaluation of the framework's capabilities in handling complex access control scenarios within EHR systems.

\subsection{Construction of policy-to-legal-ontology}
\label{subsec:ConstructionPolicyLegalOntology}
The construction of the policy-to-legal-ontology involves identifying key laws and regulations relevant to the context of electronic health records (EHR) access. For our use-case, we have focused on the legal framework within the State of Victoria in Australia, identifying three key pieces of legislation as detailed in Table \ref{tab:LegislationsEHRAccess}.

\begin{itemize}
	\item \textbf{Privacy Act 1988}\footnote{https://www.legislation.gov.au/Details/C2014C00076}: A comprehensive privacy law detailing principles around personal data collection, usage, and disclosure.
	\item \textbf{My Health Records Act 2012}\footnote{https://www.legislation.gov.au/Details/C2021C00475}: Establishes the My Health Record system, a national EHR system.
	\item \textbf{Health Records Act 2001}\footnote{https://www.legislation.vic.gov.au/in-force/acts/health-records-act-2001/047}: Defines patients' rights for health records access and health care providers' responsibilities.
\end{itemize}

\begin{table}[t!]
	\centering
	\normalsize
	\caption{List of legislations governing EHR access}
	\label{tab:LegislationsEHRAccess}
	\resizebox{\columnwidth}{!}{%
		\begin{tabular}{ccc}
			\hline
			\textbf{Legislation} & \textbf{Jurisdiction level} & \textbf{Current Version} \\ \hline
			Privacy Act 1988 & Federal & 1 Sep 2021 \\ 
			My Health Records Act 2012 & Federal & 1 Sep 2021 \\ 
			Health Records Act 2001 & State of Victoria & 2 Sep 2022 \\ \hline
		\end{tabular}%
	}
\end{table}

We incorporated the legislations into our model using the \enquote{AskYourPDF}\footnote{https://askyourpdf.com/upload} plugin of ChatGPT-4, which facilitated the importation of published PDF versions of the legislation. We did not create an explicit, clear-cut ontology model, which often proves too rigid and fails to capture the complex reality of healthcare scenarios comprehensively. Instead, we leveraged ChatGPT-4's ability to understand and retain the implications of the legislation, effectively embedding an implicit legal medical ontology within the model's attention and knowledge layers. This methodology, although unconventional, leverages the inherent flexibility of the GPT architecture, harnessing the strengths of both explicit and implicit knowledge representation. Our approach was demonstrated as a proof-of-concept implementation on ChatGPT-4, utilizing its robust hardware and computing capabilities. The resulting implicit legal medical ontology, validated under human oversight, forms the cornerstone of our GPT-Onto-CAABAC model and serves as the initial step towards our ultimate goal of creating a domain-specific \textit{Large Language Model} (LLM) trained on this ontology.

\subsection{Utilization of datasets}
\label{subsec:EmploymentDatasets}
Our strategic approach involved the construction of a comprehensive dataset comprising over 120 use-case scenarios across 12 categories to enhance the precision and reliability of GPT responses. This methodology has been indispensable for multiple reasons:

\begin{itemize}
	\item \textbf{Diverse Dataset:} Incorporating various EHR-related scenarios diversified the dataset, enriching the GPT learning experience. This diversity facilitated the model in generalizing and making accurate predictions in real-world applications.
	\item \textbf{Comprehensive Coverage:} By curating a minimum of 10 specific use-case scenarios for each category, the dataset provided an extensive representation of potential healthcare sector interactions, capturing its inherent complexities.
	\item \textbf{Cross-Referencing Legal Frameworks:} We cross-referenced the scenarios with the Australian \enquote{Privacy Act 1988} and \enquote{My Health Records Act 2012}, enabling GPT to grasp the legal consequences of various situations, thus augmenting its capacity for legally compliant recommendations.
	\item \textbf{Enhanced Accuracy:} Leveraging a large, diverse dataset fostered improvement in the GPT's responses' accuracy by exposing it to a wide range of situations and subtle contexts.
	\item \textbf{Improved Experimental Process:} Employing an expansive dataset enriched the experimental process, offering a vast source of data for training, testing, and validation, thereby strengthening the GPT model.
\end{itemize}

In our experiment, we utilized a combination of two datasets that served distinct purposes. The first dataset included anonymized, real-world EHR data, providing our system with realistic data points. The second dataset consisted of carefully constructed artificial scenarios that targeted specific capabilities of the GPT-Onto-CAABAC framework. These scenarios, which incorporated instances of high-frequency access requests, complex contextual conditions, abrupt legal or policy changes, and conflicting policies or extraordinary medical situations, offered an opportunity to evaluate the framework's robustness and adaptability. The construction of this comprehensive dataset, which included 120 use-case scenarios across 12 categories, was instrumental in addressing concerns regarding the provision of practical examples and empirical data. This dataset played a pivotal role in refining the accuracy, reliability, and legal compliance of the GPT responses. The diversity of the dataset not only facilitated the model in making accurate predictions and generalizing across various scenarios, but it also enhanced its versatility. Moreover, the alignment of the scenarios with the Australian \enquote{Privacy Act 1988} and \enquote{My Health Records Act 2012} guaranteed the model's ability to provide legally compliant recommendations. The incorporation of real-world EHR data and the tailored artificial scenarios were critical in assessing the model's adaptability and robustness under diverse conditions, yielding invaluable insights into its performance. Consequently, our methodology provided a wealth of empirical data and practical instances, highlighting the GPT-Onto-CAABAC framework's versatility, adaptability, and legal compliance. In sum, the carefully constructed dataset and the testing scenarios facilitated a rigorous examination of the model's performance, validating its potential for practical applications in healthcare access control.

\subsection{Acquiring decisions and recommendations}
\label{subsec:ObtainingDecisionsRecommendations}
The GPT-Onto-CAABAC framework employs ChatGPT-4's advanced NLP capabilities to derive access control decisions and provide recommendations. These decisions and recommendations are contingent upon two primary elements: non-negotiable policy-to-legal-ontology and negotiable context and attribute information. Both elements influence the model's understanding of EHR access control scenarios and guide its decision-making process. The non-negotiable policy-to-legal-ontology, founded on existing legal regulations and healthcare policies, constitutes a rigid baseline for decision-making. It is indispensable in ensuring adherence to pre-established privacy and security requirements in EHR data management. In this proof-of-concept stage, several strategic decisions are adopted for both practicality and exploratory value. Firstly, ChatGPT-4 is utilized in its commercial form, negating the need for retraining or fine-tuning. This decision allows for an assessment of the model's capabilities in a generic setting and offers future implementers the latitude to add domain-specific optimizations. Secondly, the framework does not retain CAABAC information but rather acquires it ad hoc for each evaluation. Such a design aligns well with the healthcare sector's inherently dynamic and complex environment, enabling adaptive access control decisions based on real-time situations rather than rigid processes. Lastly, we deliberately abstain from optimizing the model's response time at this stage. This leaves room for prospective organizations to make performance-based adjustments tailored to their specific requirements when scaling from a proof-of-concept to a full-fledged implementation.

The negotiable context and attribute information imparts the system the flexibility to adapt and respond to the dynamic, multifaceted nature of the healthcare sector. The model processes an access request by receiving a prompt describing the scenario in natural language. This prompt serves as the interface through which the context and attribute information are encoded and absorbed by ChatGPT-4. For example, a typical prompt might state: 
\begin{verbatim}
	Request for patient John Doe's EHR 
	for a clinical study by Dr. 
\end{verbatim}
\begin{verbatim}
	John Smith, who has a security clearance.
	Is access granted?
\end{verbatim}

Outputs based on such prompts could be categorized as follows:
\begin{itemize}
	\item Access granted: ``Access granted. Ensure to maintain data confidentiality.''
	\item Access denied: ``Access denied. This is illegal.''
	\item Recommendations: ``Need to seek patient's informed consent. Seek permission from ethics committee for special ethics approval.''
\end{itemize}

The model cross-checks this information against the embedded policy-to-legal-ontology. The decision is influenced not just by this ontology but also by the specific context and attributes presented, thus utilizing a form of deductive reasoning. In instances where access is denied, the model proposes recommendations for altering the context or attribute information to facilitate potential access approval. These could range from seeking permissions from higher authority to modifying the timing or environment of access. Thus, the GPT-Onto-CAABAC framework effectively balances regulatory adherence with the necessary flexibility in navigating the healthcare sector's complex landscape.

\subsection{Human evaluation and sign-off}
\label{subsec:HumanEvaluationMetrics}
The results are presented for human evaluation and sign-off. During our evalution, there is no need to sign off other than human inspection and oversight to evaluate the effectiveness of GPT decisions and recommendations. For evaluation, we need to establish quantitative metrics. These could include:

\subsubsection{Compliance} 
Measures the rate at which the system's decisions align with existing rules and policies. This could be calculated by identifying instances where the system's decisions were compliant with the rules and policies divided by the total number of decisions made. For example, if in 100 decisions, 95 were compliant with the policies, the compliance rate would be 95\%.
\subsubsection{Adaptability} 
Calculates how quickly the system adapts to sudden changes in policies or rules. This would ideally be measured over a period of time following the implementation of new rules or policies. You would compare the system's performance (in terms of compliance rate, efficiency, and recommendation quality) immediately after the change and after a certain period, say, one month. The adaptability score could be the rate of improvement in the system's performance over this period.
\subsubsection{Conflict Resolution Efficiency}
Evaluates how effectively the system resolves conflicts between different policies or rules. This could be determined by identifying cases where there was a conflict between policies or rules and seeing how often the system made the correct decision. If there were 50 cases of conflict, and the system resolved 40 correctly, the conflict resolution efficiency would be 80\%.

\subsubsection{Recommendation Quality}
The evaluation of recommendation quality requires a detailed analysis of the proposed framework's competence in capturing and interpreting Ontology and CAABAC information. This proficiency is paramount in enabling the GPT to render appropriate access control decisions. For a comprehensive examination of the GPT responses, we introduce two inherently connected key criteria: (1) \textit{Context Comprehension}, representing the system's capability to fully absorb and understand the Ontology and CAABAC information pertinent to the situation at hand, and (2) \textit{Recommendation Effectiveness}, assessing the beneficial nature and practicability of GPT's recommendations. The valuable recommendations generated by the GPT rely on its effective comprehension of the provided contextual information. Consequently, a failure in \textit{Context Comprehension} (score below 0.25) immediately translates to zero score in \textit{Recommendation Effectiveness}. We propose a ``marking rubric'' to assess system responses, mirroring a grading scheme similar to those utilized for student assignments. This rubric, outlined in Table~\ref{tab:marking_rubric}, permits the evaluation of each question against both criteria, yielding scores ranging from 0 to 1. Accordingly, a set of 10 questions can achieve a total score ranging between 0 and 10.

\begin{table*}[th!]
	\centering
	\caption{Marking rubric for evaluating GPT responses}
	\label{tab:marking_rubric}
	\begin{tabularx}{\textwidth}{|p{2cm}|p{2cm}|X|}
		\hline
		\textbf{Criteria} & \textbf{Potential Scores} & \textbf{Interpretation} \\
		\hline
		Context Comprehension & 0 - 0.5 & 0: System fails to capture the Ontology and CAABAC information in the evaluated situation. \\
		& & 0.25: System partially captures the Ontology and CAABAC information in the evaluated situation. \\
		& & 0.5: System fully captures the Ontology and CAABAC information in the evaluated situation. \\
		\hline
		Recommen-dation Effec-tiveness & 0 - 0.5 & 0: GPT's recommendations are not beneficial, require extensive human improvements, or if Context Comprehension score is 0. \\
		& & 0.25: GPT's recommendations are somewhat beneficial, and require moderate human improvements.\\
		& & 0.5: GPT's recommendations are highly beneficial and require little to no improvements.\\
		\hline
	\end{tabularx}
	
\end{table*}

\section{Evaluations}
\label{sec:evaluation_insights}
Our evaluation of the GPT-Onto-CAABAC framework extends beyond traditional metrics, delving into the nuanced capabilities of GPT-powered access control within the intricate landscape of EHR security. This rigorous analysis unveils the framework's innovative approach to dynamically interpreting access control policies, showcasing its adaptability and compliance with existing healthcare regulations.

\subsection{Scenario Testing with Evaluation Metrics}
\label{subsec:ModelSimulationScenarioTesting}
The GPT-Onto-CAABAC framework underwent comprehensive scenario testing, reflecting real-world healthcare decision-making complexities. These scenarios, rigorously designed to assess the framework's proficiency in navigating hospital policies, legal requirements, and patient-specific contexts, also scrutinized its adaptability across various roles. The evaluation emphasized not only the role-based and patient-consent-driven access but also the framework's resilience through fault injection testing, highlighting its robustness and adaptability in managing complex, dynamic scenarios.

\subsubsection{Scenario testing}
\label{subsubsec:ScenarioTesting}
Through meticulous scenario testing, we explored the GPT model's adeptness in interpreting EHR access control's legalities and ethics. The model's ability to comprehend context and provide actionable recommendations was particularly noteworthy, demonstrating its potential as a decision-support tool in healthcare. This testing phase illustrated the model's superior grasp of privacy laws and healthcare protocols, showcasing its nuanced understanding of role-specific permissions and the critical importance of patient consent.

\subsubsection{Fault injection testing}
\label{subsubsec:FaultInjectionTesting}
The fault injection testing phase offered insights into the GPT-Onto-CAABAC framework's ability to navigate misleading situations, further affirming its competency in handling ethically and legally complex scenarios. The model's performance, evaluated against the backdrop of the My Health Records Act 2012, was commendable, with its recommendations aligning closely with human expectations and legal standards. This phase underlined the framework's promise in augmenting medical access control risk auditing, suggesting its utility in identifying and rectifying potential compliance deviations.

The GPT-Onto-CAABAC framework distinguishes itself by offering a policy-compliant spectrum of options, echoing the need for a flexible, human-centric approach in interpreting dynamic policies. This nuanced capability, set against the rigidity of traditional access control systems, underscores the potential of GPT-powered frameworks to revolutionize EHR security by infusing adaptability and intelligence into access control decisions.

\begin{figure*}[tp!]
	\centering
	\begin{tikzpicture}
	\end{tikzpicture}
	\caption{Evaluation of GPT Answers Per Category (higher is better)}
	\label{fig:EvaluationGPT}
\end{figure*}

\subsubsection{Enhancing Conflict Resolution with GPT}
\label{subsec:enhancingConflictResolutionGPT}
This study delineates the efficacy of conflict resolution mechanisms utilizing the Generative Pre-trained Transformer (GPT) model within the GPT-Onto-CAABAC framework. The GPT model, with its advanced natural language processing capabilities, dynamically processes access requests by comprehensively analyzing them against a backdrop of established policies alongside user-specified attributes. This innovative approach not only facilitates the identification of potential conflicts but also recommends resolutions. These resolutions are deeply rooted in the integrated ontologies of medical and legal domains and are further refined by the Context-Aware Attribute-Based Access Control (CAABAC) system's context-sensitive parameters.

Such a mechanism ensures that every decision upholds the highest standards of regulatory compliance while being intricately customized to the particularities of the request's context. This flexible and customized access control method marks a significant advancement in navigating complex and ever-changing healthcare environments. Using this framework, healthcare providers can achieve a balance between stringent security measures and the need for adaptive, context-aware access to sensitive information.
\cite{aydin2021,shadish2021,yao2021,gupta2020,ando2022,wang2020,pereira2021,yuan2021,zhang2020privacy}.

\subsection{Future Directions for Comprehensive Evaluation}
\label{subsec:FutureDirectionsComprehensiveEvaluation}
The study recognizes an imperative for a more expansive evaluation of the GPT-Onto-CAABAC framework. To this end, a forward-looking agenda for augmented experiments and analysis is proposed, aimed at thoroughly validating the framework's performance and applicability in real-world contexts.

\textbf{Acknowledging GPT's Unique Capabilities:} It is crucial to underline that GPT-powered access control systems diverge significantly from traditional models in their operational philosophy. Unlike the rigid, machine-dictated approaches of conventional systems, GPT-based frameworks excel in interpreting and adapting to highly dynamic policies, infusing a level of flexibility and human-like understanding previously unattainable. This inherent difference necessitates a unique evaluation perspective, one that appreciates the qualitative enhancements that GPT introduces to access control, from interpreting complex scenarios to advising on compliance in ways traditional systems cannot.

\textbf{Extended Scenario Testing:} Future experiments will broaden scenario testing to encompass a diverse range of healthcare contexts, with the aim of capturing the adaptability and efficacy of the GPT-Onto-CAABAC framework in various operational scenarios.

\textbf{Quantitative Performance Metrics:} We will complement qualitative insights with quantitative metrics, such as accuracy, response time, and fault tolerance, offering a balanced view of the performance characteristics of the framework.

\textbf{Real-World Pilot Studies:} Implementing pilot studies within actual healthcare environments will bridge the gap between theoretical assessment and practical application, providing a direct insight into the real-world utility of the framework and areas for improvement.

\textbf{User Feedback and Iterative Refinement:} Gathering feedback from end-users and subject matter experts will be paramount. This iterative process will ensure the framework's evolution in alignment with user expectations and industry standards, refining its functionality and user experience.

\textbf{Comparative Analysis:} A comparative analysis with traditional access control models will highlight the GPT-Onto-CAABAC framework's novel capabilities, particularly its adaptability and intelligent decision-making, illustrating a significant leap over the limitations of conventional access control systems.

This comprehensive approach to future evaluation endeavors not only addresses the reviewer's concerns but also emphasizes the paradigm shift GPT-powered access control represents in managing EHR security. By advancing these efforts, we aim to substantiate the framework's transformative potential and its alignment with the evolving landscape of healthcare data management.

Our analysis reaffirms the GPT-Onto-CAABAC framework's innovative intersection with AI and healthcare regulation, spotlighting its capacity to enhance the privacy and security of EHR systems. While the framework's current evaluation highlights significant strides in AI-powered access control, ongoing refinement and real-world testing remain imperative to fully realize its transformative potential in healthcare data management.

\subsubsection{GPT responses patterns}
\label{subsubsec:GPTResponsesPatterns}
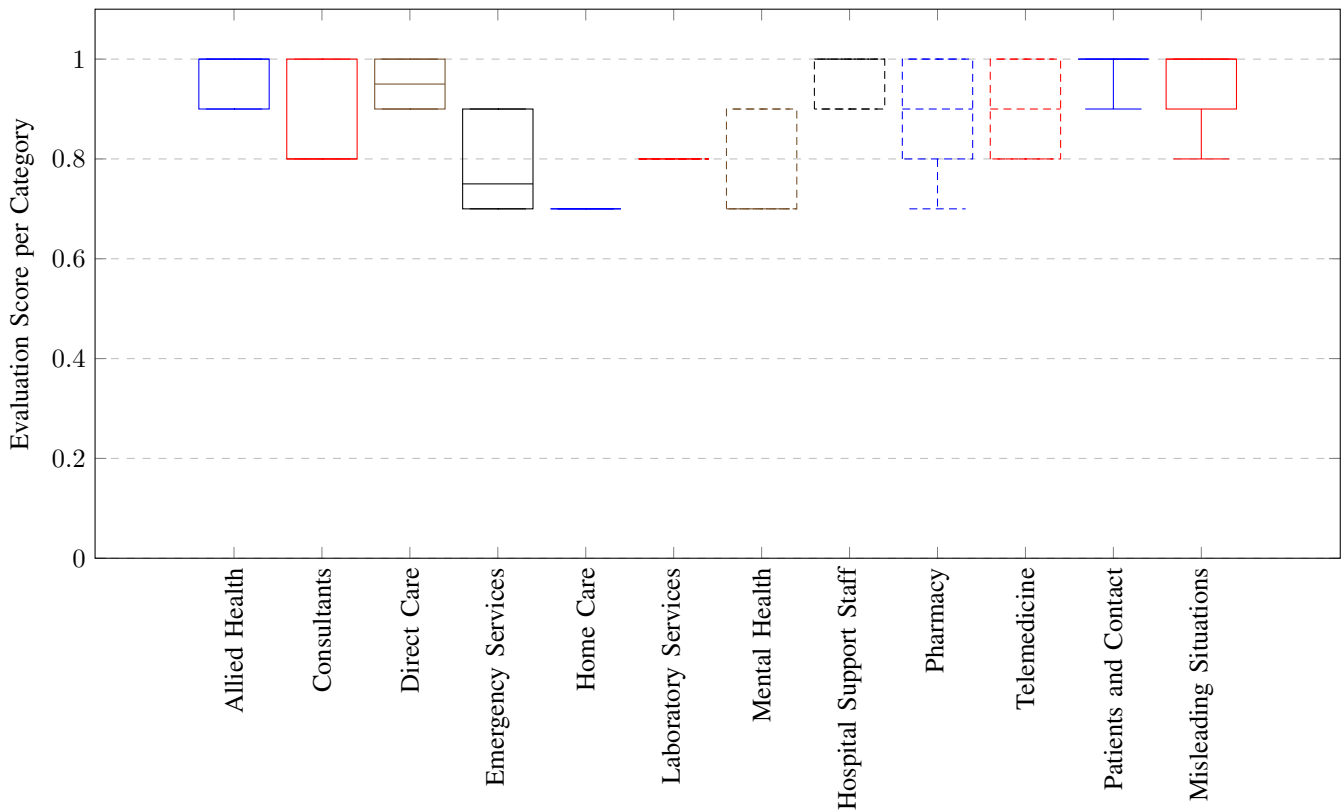
\begin{figure*}[tp!]
	\normalsize
	\centering
	\begin{tikzpicture}
		\begin{axis}[
			boxplot/draw direction=y,
			xtick={1,2,3,4,5,6,7,8,9,10,11,12},
			ymin=0,
			ymajorgrids=true,
			grid style=dashed,
			ytick={0,0.2,...,1.0},
			xticklabels={Allied Health,Consultants,Direct Care,Emergency Services,Home Care,Laboratory Services,Mental Health,Hospital Support Staff,Pharmacy,Telemedicine,Patients and Contact,Misleading Situations},
			x tick label style={rotate=90,anchor=east},
			ylabel={Evaluation Score per Category},
			width=\textwidth,
			height=0.49*\textwidth,
			]
			
			\addplot+[boxplot prepared={lower whisker=0.9, lower quartile=0.9, median=1, upper quartile=1, upper whisker=1}] coordinates {};
			
			\addplot+[boxplot prepared={lower whisker=0.8, lower quartile=0.8, median=0.8, upper quartile=1, upper whisker=1}] coordinates {};
			
			\addplot+[boxplot prepared={lower whisker=0.9, lower quartile=0.9, median=0.95, upper quartile=1, upper whisker=1}] coordinates {};
			
			\addplot+[boxplot prepared={lower whisker=0.7, lower quartile=0.7, median=0.75, upper quartile=0.9, upper whisker=0.9}] coordinates {};
			
			\addplot+[boxplot prepared={lower whisker=0.7, lower quartile=0.7, median=0.7, upper quartile=0.7, upper whisker=0.7}] coordinates {};
			
			\addplot+[boxplot prepared={lower whisker=0.8, lower quartile=0.8, median=0.8, upper quartile=0.8, upper whisker=0.8}] coordinates {};
			
			\addplot+[boxplot prepared={lower whisker=0.7, lower quartile=0.7, median=0.7, upper quartile=0.9, upper whisker=0.9}] coordinates {};
			
			\addplot+[boxplot prepared={lower whisker=0.9, lower quartile=0.9, median=1, upper quartile=1, upper whisker=1}] coordinates {};
			
			\addplot+[boxplot prepared={lower whisker=0.7, lower quartile=0.8, median=0.9, upper quartile=1, upper whisker=1}] coordinates {};
			
			\addplot+[boxplot prepared={lower whisker=0.8, lower quartile=0.8, median=0.9, upper quartile=1, upper whisker=1}] coordinates {};
			
			\addplot+[boxplot prepared={lower whisker=0.9, lower quartile=1, median=1, upper quartile=1, upper whisker=1}] coordinates {};
			
			\addplot+[boxplot prepared={lower whisker=0.8, lower quartile=0.9, median=1, upper quartile=1, upper whisker=1}] coordinates {};
		\end{axis}
	\end{tikzpicture}
	\caption{Variation of Evaluation Scores of GPT Responses By Category}
	\label{fig:variations}
\end{figure*}

Our GPT-Onto-CAABAC framework, in its interpretation of legal boundaries for EHR access, demonstrates a rich and complex range of responses across different scenarios. These responses, depicted in Fig. \ref{fig:variations}, highlight the multifaceted nature of this AI system and its ability to understand and adapt to intricate contexts.

Upon an in-depth analysis of the patterns emerging from the GPT's responses, five key categories of variations were identified: role-specific permissions, policy adherence, patient consent, healthcare purpose, and supervision.

\begin{itemize}
	\item \textbf{Role-specific permissions:} As illustrated by the data, role specificity has a profound impact on GPT responses. For categories like consultants, allied health, and direct care, GPT models showed near-perfect adherence to policy. For roles with less well-defined policy boundaries, such as emergency services, mental health, and hospital support staff, a slight decrease in the evaluation score was observed. These lower scores may result from the relative ambiguity in access control policies specific to these roles, requiring more intricate judgment from the GPT model.
	
	\item \textbf{Policy adherence:} Policies outlined in the My Health Records Act 2012 form the backbone of the access control decisions. GPT models exhibited excellent comprehension of these policies, as observed in high scores across most categories. However, variations exist; in the case of misleading situations or home care, where personal relationships and less formal care settings blur the policy lines, the evaluation scores slightly dropped. This may reflect GPT's struggle to balance legal policy with complex human situations.
	
	\item \textbf{Patient consent:} Consent is a crucial factor in healthcare data access. GPT's interpretation of consent-focused scenarios received commendable scores, especially when dealing with the 'Patients and Contact' category. The slightly lower score in 'Misleading Situations' may be attributed to the ambiguity introduced by the presence of close relationships, which challenges the strict legal interpretation of patient consent.
	
	\item \textbf{Healthcare purpose:} GPT's responses accurately reflected the healthcare-centric purpose of EHR access, achieving high scores in areas such as direct care, consultants, and telemedicine. Lower scores in home care and emergency services, however, suggest the model's difficulty in discerning purpose in crisis situations or informal care environments.
	
	\item \textbf{Supervision:} In situations involving supervised roles, such as students or interns, GPT was adept at incorporating the need for oversight into its responses. The lower score for 'Laboratory Services' may suggest the need for improved model training on subtle roles that might require supervision.	
\end{itemize}

These variations offer valuable insights into the subtle performance of the GPT-Onto-CAABAC framework. The fluctuating scores across categories point to the AI's struggles and successes in interpreting complex legal and ethical issues surrounding EHR access. While GPT models excel in clearly defined situations, they show difficulty when handling ambiguous or emotionally charged contexts. Hence, while the GPT model is an impressive tool for interpreting access control decisions, these results highlight the essential need for human oversight. Variations in response patterns underline the ongoing challenge of refining AI models to comprehend the full complexity of real-life situations and indicate potential areas for future improvement. The interpretation of these variations can assist in developing more accurate, context-sensitive AI systems for the future.

\subsection{Comparative evaluation}
\label{subsec:ComparativeEvaluation}
GPT models like GPT-3 and GPT-4 have demonstrated notable competencies in understanding and generating human-like text. Their adaptability across various tasks, even without task-specific data, proves beneficial in domains such as healthcare and law, where dynamic interpretations of user roles and corresponding access rights are essential. However, their decision-making process can be time-consuming, contrasting with the immediate decisions rendered by traditional access controls based on pre-set rules and policies. In healthcare, GPT models offer extensive patient histories, suggest relevant medical tests, and assist in honing differential diagnoses. Our scenario tests (Subsection \ref{subsubsec:ScenarioTesting}) demonstrated the GPT-Onto-CAABAC framework's adept understanding of the My Health Records Act 2012, effectively handling diverse healthcare roles. Yet, its efficacy in real-world conflicts requires further exploration. GPT also shows promise in legal contexts, with abilities to interpret complex legal documents, formulate legal arguments, and even predict legal outcomes. Our fault injection tests (Subsection \ref{subsubsec:FaultInjectionTesting}) evidenced that the GPT model provided policy-compliant options even in deceptive scenarios, underscoring its robustness in interpreting legal aspects related to EHR access control decisions. 

Traditional access controls, while less adaptable to rule or policy changes and requiring manual adjustments, offer the advantage of speed in decision-making, especially in time-critical, real-time scenarios. However, GPT models adapt swiftly to new data and context shifts, providing a vital edge in settings with evolving access control needs. The extent of this adaptability, for both GPT and traditional models, largely depends on the use case specifics and system programming. Despite their slower response time, GPT models' significant benefits lie in their adaptability and flexibility. They are particularly useful for post-mortem audits in risk management, employing their capability for detailed text generation to offer valuable insights for risk assessment and mitigation. As revealed by the GPT response patterns (Subsection \ref{subsubsec:GPTResponsesPatterns}), the variable performance of GPT models under different conditions underscores the need for human oversight and suggests areas for potential improvement.

\subsection{Ethical and societal implication analysis}
\label{subsec:EthicalSocietalImplication}
In the context of EHR access control, ethical and societal implications primarily revolve around conflicts that could emerge from varying access rights associated with different roles, and potential disagreements regarding patient consent. Notably, the scenario tests conducted to evaluate the performance of the GPT-Onto-CAABAC framework did not explicitly present any such conflicts requiring resolution. Potential conflicts could, however, surface in real-world settings. These could stem from contradictions between the access permissions of distinct roles, like healthcare professionals and relatives of the patient, especially when their interests do not align. Similarly, situations might arise wherein disagreements over patient consent could trigger conflicts, posing a substantial challenge to the decision-making process.

The GPT-Onto-CAABAC framework's proficiency in addressing and resolving such conflicts can be adequately gauged only when it is confronted with actual conflict scenarios. As such, despite promising preliminary results from the initial tests, it remains crucial to subject the framework to rigorous and comprehensive testing simulating real-world conflict scenarios to fully ascertain its effectiveness and readiness for practical implementation.

\subsection{Assessment of Transparency and Interpretability}
\label{subsec:TransparencyInterpretability}
Addressing prevalent concerns around the ``black box'' phenomenon in AI systems, we made a conscious effort to evaluate the transparency and interpretability of the GPT-Onto-CAABAC framework. The primary aim was to discern whether the framework's decision-making process and outputs were intuitively understandable and accessible to healthcare professionals or policy makers. The assessment, far from being a superficial overview, entailed a thorough examination of the GPT-Onto-CAABAC framework's rationale behind EHR access control decisions. This rigorous investigation intended to ensure that healthcare professionals or policy makers could readily understand the framework's logic, thereby facilitating informed decisions regarding EHR access control based on the framework's insights.

Our framework demonstrated consistent response patterns across diverse scenarios, which substantially bolstered its interpretability. It provided satisfactory reasoning based on factors such as role-specific permissions, policy adherence, patient consent, healthcare purpose, and supervision. While processing requests and offering recommendations, it effectively accounted for various aspects defined by the My Health Records Act 2012. The analysis indicated a substantial degree of transparency and interpretability in the framework's decision-making process, increasing its potential utility in a real-world healthcare setting. Although these promising results are encouraging, continued refinement and testing of the framework's capabilities, particularly for complex scenarios, are necessary to further enhance its transparency and interpretability. Balancing this need with human oversight, especially in ambiguous or emotionally charged situations, is crucial. The GPT-Onto-CAABAC framework's transparency and interpretability assessment results demonstrated its capacity to offer decision-making processes that are comprehensive, consistent, and accessible to end-users, thereby suggesting its potential as a viable decision-support tool in healthcare settings.

\section{Discussions}
\label{sec:discussions}
This section delves into a comprehensive discussion of the significant issues that emerged during the experiment. 

\subsection{Challenges and Overcoming Strategies}
\label{subsec:challenges_gpt}
The implementation of the GPT-Onto-CAABAC framework within healthcare, despite its significant potential, presents several salient challenges. The complexity of healthcare scenarios, performance and validity issues, and the overarching concern of societal trust necessitate a systematic addressal. However, these challenges also present opportunities for further refinement and innovation.

\begin{itemize}
	\item \textbf{Stability of GPT-generated texts}: In our pilot testing, we found that GPT produces slight variations in its outputs for the same input, primarily linguistic rather than semantic. We propose regular audits and ongoing scrutiny to ensure the consistency and reliability of GPT-generated content. Additionally, implementing feedback loops from end-users can provide valuable insights for model fine-tuning.
	
	\item \textbf{Performance of the GPT models}: With the increasing sophistication and size of GPT models, there's an associated increase in response generation time, making the framework unsuitable for real-time, time-critical decision-making in healthcare. To tackle this, we recommend continued performance evaluations and the development of optimization strategies. This may involve parallel processing, model pruning, or exploring hardware acceleration options.
	
	\item \textbf{Validity of GPT-based decisions}: The potential of GPT models to produce hallucinations - factually incorrect or irrelevant outputs - could lead to non-compliant healthcare decisions \cite{mcintosh2023culturally}. To mitigate this risk, it is crucial to implement continuous validation checks and a verification mechanism. This might involve cross-checking GPT outputs with trusted resources, implementing peer-review mechanisms, or integrating GPT with rule-based systems for sanity checks.
	
	\item \textbf{Societal trust in AI systems}: The potential for hallucinations and the opaque nature of AI algorithms present a significant challenge in fostering societal trust. For this, we advocate for strong human oversight, robust mechanisms for GPT output validity monitoring, and effective public communication strategies. Transparency about model limitations, clear communication about how decisions are made, and maintaining accountability are essential in earning public trust. Additionally, collaboration with regulatory bodies and ethicists to design guidelines and policy frameworks can contribute to societal trust.
\end{itemize}

Addressing these challenges isn't a one-time activity but requires an ongoing cycle of refining and evaluating the GPT-Onto-CAABAC framework. It is through continuous iteration that we can enhance performance, validate results, improve transparency, and maintain effective public communication to harness the power of this framework in healthcare decision-making.

\subsection{Applications in Healthcare Settings}
\label{subsec:utilization}
Our GPT-Onto-CAABAC framework offers an adaptable solution fitting a variety of healthcare scenarios. Its flexibility facilitates its employment across healthcare decision-making domains, acting either as a proactive recommendation system or a reactive risk management tool. Traditional security consultations in healthcare are plagued by challenges such as the intensive manual work required for auditing intricate policies, unclear interpretations of regulations, and the rigidity to adjust to new policies. These issues, combined with often inadequate insights, could affect the efficacy of consultations. The GPT-Onto-CAABAC framework confronts these challenges head-on. LLMs automate auditing, drastically reducing manual involvement. GPT models' natural language skills clarify complex healthcare contexts, and the framework's continual learning feature keeps it aligned with shifting regulations. This combined prowess offers healthcare professionals a reliable decision-making aid.

Proactively, our framework guides early decision-making stages, presenting policy-aligned alternatives for intricate clinical situations. Here, GPT models comprehend detailed patient data, while ontology systems furnish context-driven advice based on policy and regulatory interpretations. This cohesive method promotes complex decision-making tailored to each case's specifics. As a reactive mechanism, the GPT-Onto-CAABAC system reviews healthcare decisions post-facto, ensuring they adhere to legal and organizational standards while spotlighting non-conformities. This retrospective review ensures consistent policy adherence, highlights training needs, and pinpoints policy areas needing further clarity. Additionally, this framework holds potential as an educational asset in healthcare training. Through the analysis of prior decisions, it can refine academic syllabi, shedding light on the intricate relationship between healthcare methods, policy mandates, and genuine patient situations. Despite its evident value, it remains essential to evaluate the GPT-Onto-CAABAC framework's effectiveness across diverse healthcare settings, ensuring its continued relevance and contribution to healthcare decision processes.

\subsection{Integrating Human Oversight in the GPT-Onto-CAABAC Framework}
\label{subsec:HumanOversightIntegration}

The study addresses the Human Oversigh within the GPT-Onto-CAABAC framework, and we emphasize its critical role in enhancing the decision-making process. Although GPT and the CAABAC model offer robust automated capabilities for access control and conflict resolution, human oversight serves as an essential layer to ensure ethical compliance, accountability, and adaptability to complex scenarios that automated systems might not fully grasp. This integration allows for a comprehensive review of automated decisions, particularly in sensitive cases, ensuring that they are in accordance with organizational policies, legal standards, and ethical considerations. Through this collaborative approach, the framework not only leverages the efficiency of automation but also retains the discerning judgment of human experts, effectively resolving the potential limitations of relying solely on automated processes \cite{aydin2021,shadish2021,yao2021,gupta2020,ando2022}.

\subsection{Expanded use cases beyond EHR}
\label{subsec:UseCases}
Our GPT-Onto-CAABAC framework has broad applicability across diverse sectors that require complex and detailed access control decisions considering compliance, context, and attributes. Below are some potential use cases:

\begin{itemize}
	\item \textbf{Financial Services:} In the financial sector, access controls for sensitive customer data need to balance privacy regulations, individual access needs, and security priorities. The framework can aid in compliant access control by considering financial advisor attributes, customer consent context, and privacy laws.
	
	\item \textbf{Defense Organizations:} For defense organizations, granting access to classified data requires strict adherence to security protocols and hierarchies. The framework can incorporate user roles, context like emergency situations, and classification levels to make informed yet flexible access decisions.
	
	\item \textbf{Legal Services:} In legal services, client confidentiality is paramount while collaborating with experts across specializations. The framework can weigh attorney attributes, client permissions, and legal ethics codes to enable secure yet productive information sharing.
	
	\item \textbf{Public Sector:} Government agencies manage vast sensitive citizen data subject to complex regulations. The framework can help navigate user clearances, data types, compliance needs, and transparency laws for responsible public data access.
	
	\item \textbf{Research Institutions:} Academic research requires collaborations across domains while protecting participant privacy. The framework can balance researcher credentials, study protocols, ethics approvals and privacy laws to uphold rigorous access control standards.
\end{itemize}

\begin{figure*}[t!]
	\centering
	\includegraphics[width=\textwidth]{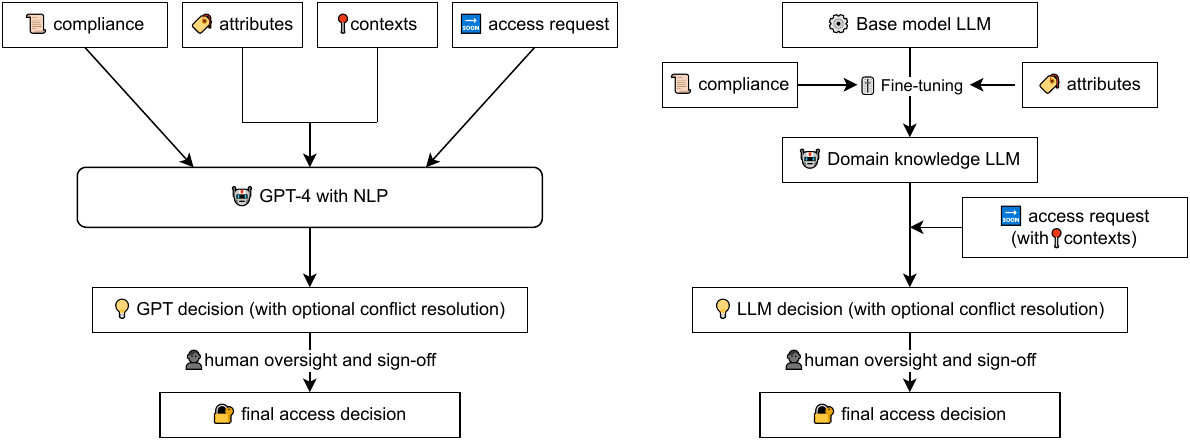}
	\caption{Comparison of our GPT-4-based prototype (left) and a practical domain knowledge LLM implementation (right)}
	\label{fig:comparision}
\end{figure*}

\subsection{Translating concept to real-world implementation}
While the GPT-Onto-CAABAC framework shows promise as a conceptual model, translating it into large-scale healthcare implementation requires the adoption of a fine-tuned domain knowledge LLM (Fig. \ref{fig:comparision}), and necessitates significant translational research and stakeholder engagement. Some key aspects should be considered:

\begin{itemize}
	\item \textbf{Pilot Testing and Optimization}: Extensive testing across diverse healthcare contexts, institutions and geographic regions is crucial. This allows for framework optimization and customization based on lessons learned during deployment.
	
	\item \textbf{Regulatory Approvals}: Securing approvals from healthcare governance bodies and demonstrating compliance is pivotal prior to full-scale rollout. This ensures patient safety and security standards are met.
	
	\item \textbf{Change Management}: Training healthcare professionals on integrating the framework into workflows is vital. Managing organizational change and addressing adoption barriers smooths the transition.
	
	\item \textbf{Patient Advocacy}: Incorporating patient perspectives through focus groups and consultation can identify potential ethical concerns early. Their insights further bolster framework transparency.
	
	\item \textbf{Continuous Improvement}: Updating the framework as healthcare regulations and AI advance is imperative. Establishing processes for regular enhancements sustains long-term relevance.
	
	\item \textbf{Economic Analysis}: Conducting cost-benefit analysis guides budgeting and resource allocation for development and maintenance. Quantifying value gained aids wider adoption.
\end{itemize}

\begin{figure}[t!]
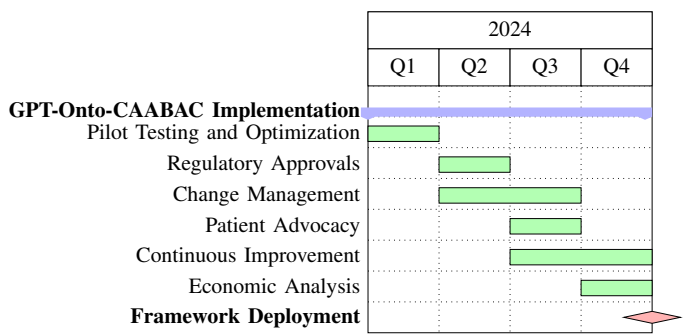

	\centering
	\resizebox{\columnwidth}{!}{ 
		\begin{ganttchart}[
			hgrid,
			vgrid,
			x unit=1.15cm,
			y unit title=0.6cm,
			y unit chart=0.5cm,
			title height=1,
			bar height=0.5,
			group right shift=0,
			group top shift=0.7,
			group height=.3,
			group peaks width={0.2},
			milestone label font=\bfseries\normalsize,
			bar label font=\normalsize,
			title label font=\normalsize,
			group/.append style={fill=blue!30},         
			bar/.append style={fill=green!30},          
			milestone/.append style={fill=red!30}       
			]{1}{4}
			
			\gantttitle{2024}{4} \\
			\gantttitle{Q1}{1} \gantttitle{Q2}{1} \gantttitle{Q3}{1} \gantttitle{Q4}{1} \\
			
			\ganttgroup{GPT-Onto-CAABAC Implementation}{1}{4} \\
			\ganttbar{Pilot Testing and Optimization}{1}{1} \\
			\ganttbar{Regulatory Approvals}{2}{2} \\
			\ganttbar{Change Management}{2}{3} \\
			\ganttbar{Patient Advocacy}{3}{3} \\
			\ganttbar{Continuous Improvement}{3}{4} \\
			\ganttbar{Economic Analysis}{4}{4} \\
			\ganttmilestone{Framework Deployment}{4} 
			
		\end{ganttchart}
	}
	\caption{GPT-Onto-CAABAC Implementation Roadmap for 2024}
	\label{fig:gantt}
\end{figure}

The Gantt chart, shown in Figure \ref{fig:gantt}, visualizes the implementation timeline for 2024. The chart has been derived based on expert estimates and stakeholder inputs:

\begin{itemize}
	\item \textbf{Pilot Testing and Optimization} is scheduled for Q1, considering it is the primary phase to validate the framework.
	\item \textbf{Regulatory Approvals} are set in Q2, once preliminary results from pilot tests are available.
	\item \textbf{Change Management} spans from Q2 to Q3, as training and transition management processes often overlap with other tasks.
	\item \textbf{Patient Advocacy} is planned for Q3, ensuring ethical considerations are reviewed and integrated.
	\item \textbf{Continuous Improvement} begins from Q3 and extends to Q4, emphasizing ongoing updates based on the framework's deployment feedback.
	\item \textbf{Economic Analysis} is conducted in Q4 to guide further resource allocation and budgeting decisions.
\end{itemize}

This phased translational approach is key to overcoming operational complexities and bridging the gap from conceptual model to field deployment. With diligent pilot testing, stakeholder engagement, iterative improvements and economic prudence, the GPT-Onto-CAABAC framework can progress from theory to practice.

\subsection{Comparative Analysis and Evaluation of GPT-Onto-CAABAC}
The study extends the discussion to include a comparative analysis of the GPT-Onto-CAABAC framework against existing solutions. This analysis not only highlights our framework's unique contributions but also situates it within the broader landscape of AI and healthcare access control.

\begin{itemize}
	\item \textbf{Adaptability and Context-Awareness:} Unlike traditional access control systems, the GPT-Onto-CAABAC framework offers superior adaptability and context-awareness, crucial for dynamic healthcare environments. Our framework's use of GPT models and ontologies allows for a nuanced understanding of user roles and access needs in real-time, outperforming conventional systems that often require manual updates \cite{gupta2020,abd2021access,kayes2015ontcaac}.
	
	\item \textbf{Real-World Application and Scalability:} Through scenario-based evaluations, we demonstrate the GPT-Onto-CAABAC framework's practical applicability and scalability. Compared to existing models, which are typically validated in controlled or small-scale settings, our approach is tested against a variety of complex healthcare scenarios, showcasing its readiness for broader implementation \cite{1874376:27758905,1874376:27758933, 1874376:27758956, basil2022health}.
	
	\item \textbf{Interpretability and Transparency:} The framework improves upon the opacity often associated with AI systems. By integrating ontological reasoning with GPT's natural language processing capabilities, it offers interpretability and transparency in decision-making, a step forward from the "black box" nature of many AI tools \cite{ca_abac_integration,fragidis2018implementation}.
	
	\item \textbf{Security and Privacy:} Security and privacy considerations are paramount in our framework. Using attribute-based access control enriched with contextual data, GPT-Onto-CAABAC ensures that access decisions meet the highest standards of data protection, compared favorably with existing solutions that may not offer the same level of granular control \cite{nowrozy2023towards,huang2019ontology,dhanapal2021secure}.
	
	\item \textbf{Integration with Existing Systems:} Our framework is designed for compatibility with existing EHR systems, facilitating seamless integration. This aspect is particularly noteworthy when compared to other proposals that may necessitate substantial modifications to current infrastructures \cite{DPHealthcare2021,el2020survey,ethical_decision_paper,coorevits2013electronic,carter2022electronic}.
	
	\item \textbf{Future-Proofing and Flexibility:} Finally, the GPT-Onto-CAABAC framework is built with future developments in mind. Its modular design allows for easy updates as AI technology and healthcare practices evolve, offering a flexible solution that remains relevant over time \cite{1874376:27758905,nowrozy2023towards,wang2018towards,bhuyan2021role}.
\end{itemize}

This comparative analysis underscores the innovative contributions of the GPT-Onto-CAABAC framework and its potential to address the current and future challenges in healthcare access control.

\subsection{Research limitations}
\label{subsec:limitations}
Our research presented in article primarily focuses on the application of GPT models, ontology systems, and CAABAC models in the context of EHR access control. Some potential limitations of Our research could include:

\begin{itemize}
	\item The research might be limited by the quality and quantity of the data used for training the GPT models. If the data is not diverse or comprehensive enough, the models might not perform optimally in real-world scenarios.
	\item The research might also be limited by the complexity of integrating multiple systems (GPT models, ontology systems, and access control models). This integration might present challenges in terms of system compatibility, data synchronization, and performance optimization.
	\item The research might be limited by the rapidly evolving nature of both healthcare regulations and AI technologies. The proposed framework might need to be continuously updated to keep up with these changes.
\end{itemize}

\subsection{Future research directions}
\label{subsec:future}
Given the potential limitations of our study, we believe the future research could focus on:

\begin{itemize}
	\item Improving the quality and diversity of the training data for the GPT models. This could involve collecting more data from a wider range of sources, or developing new data augmentation techniques.
	\item Converting the framework into a domain-knowledge LLM tailored for specific use cases, as detailed in Section \ref{subsec:UseCases}.
	\item Exploring more efficient ways to integrate the GPT models, ontology systems, and access control models. This could involve developing new algorithms or system architectures.
	\item Keeping up with the latest developments in healthcare regulations and AI technologies. This could involve regular literature reviews, or collaborations with regulatory bodies and AI research institutions.
\end{itemize}

\section{Conclusions}
\label{sec:conclusions}
Our proposed GPT-Onto-CAABAC framework has advanced EHR access control by incorporating advanced AI capabilities, presenting a dynamic, context-aware model. This integration has the potential to revolutionize healthcare data security and address the multifaceted complexities of EHR access control comprehensively. The ontology-driven component provides a structured methodology for defining crucial concepts such as users, resources, roles, permissions, and contextual data, underpinning coherent access policy articulation, thereby strengthening EHR security. The system's adaptability is augmented through CAAC and ABAC integration, enhancing its applicability across varied healthcare contexts. With the GPT model's inclusion, the system can leverage sophisticated NLP capabilities, facilitating extraction and interpretation of complex legal and regulatory data, thereby enriching decision-making processes. The design of our model promotes adaptability and efficiency while upholding accountability principles, with inbuilt mechanisms for human evaluation and oversight to foster responsible AI use. Nevertheless, GPT-Onto-CAABAC's deployment is not without challenges. Effective implementation requires substantial resources and expertise, potentially challenging for smaller healthcare entities. Moreover, given the rapid evolution of healthcare and technology, the model requires regular updates to maintain its relevance. It is essential to balance potential conflicts between ontology, CAAC, and ABAC components and manage the ethical implications of deployment, particularly given the sensitive nature of EHR data. 

Beyond its immediate application in healthcare, the proposed model demonstrates considerable promise for broader implications. The model's inherent design showcases immense potential for auditing access control decisions not only in healthcare but across varied sectors. Industries with multidimensional policies, rapidly changing contexts, and the necessity for detailed post-decision audits could significantly benefit from such a model. This opens avenues for the GPT-Onto-CAABAC framework to elevate access control auditing across many critical and dynamic environments. Despite these hurdles and the expanded potential of the model, our GPT-Onto-CAABAC framework represents a significant advancement towards integrating state-of-the-art AI capabilities into EHR access control. The dynamism, adaptability, robustness, and context-aware attributes of the model enable it to meet evolving healthcare demands while adhering to prevailing regulations and policies. Future research should focus on optimizing GPT model training data, refining the integration of GPT models, ontology systems, and access control models, and staying abreast of healthcare regulations and AI technologies. As the field progresses, we anticipate the GPT-Onto-CAABAC model will continue to be a novel and adaptable solution, enhancing its efficacy across diverse healthcare scenarios, and pushing the boundaries of AI application in healthcare.

In summary, while the GPT-Onto-CAABAC framework introduces significant advancements in EHR access control, its reliance on advanced AI and ontology models introduces complexities in implementation and requires substantial computational resources. Future research could explore optimizing these aspects to enhance scalability and reduce overhead. Additionally, the evolving nature of GPT models requires continuous monitoring for ethical and privacy implications, suggesting further refinement in incorporating robust ethical guidelines and privacy-preserving mechanisms.

\section*{Abbreviations}
The table shows a list of abbreviations used in this article.
\begin{table}[h!]
	\label{tab:abbreviations}
	\begin{tabular}{ll}
		ABAC & Attribute-Based Access Control \\
		AI & Artificial Intelligence \\
		CAABAC & Context-Aware Attribute-Based Access Control \\
		CAAC & Context-Aware Access Control \\
		EHR & Electronic Health Record \\
		GPT & Generative Pre-trained Transformers \\
		LLM & Large Language Model \\
		NLP & Natural Language Processing \\
		RBAC & Role-Based Access Control
	\end{tabular}%
\end{table}

\bibliographystyle{IEEEtran} 
\bibliography{references}

\begin{thebibliography}{100}
\providecommand{\url}[1]{#1}
\csname url@samestyle\endcsname
\providecommand{\newblock}{\relax}
\providecommand{\bibinfo}[2]{#2}
\providecommand{\BIBentrySTDinterwordspacing}{\spaceskip=0pt\relax}
\providecommand{\BIBentryALTinterwordstretchfactor}{4}
\providecommand{\BIBentryALTinterwordspacing}{\spaceskip=\fontdimen2\font plus
\BIBentryALTinterwordstretchfactor\fontdimen3\font minus
  \fontdimen4\font\relax}
\providecommand{\BIBforeignlanguage}[2]{{%
\expandafter\ifx\csname l@#1\endcsname\relax
\typeout{** WARNING: IEEEtran.bst: No hyphenation pattern has been}%
\typeout{** loaded for the language `#1'. Using the pattern for}%
\typeout{** the default language instead.}%
\else
\language=\csname l@#1\endcsname
\fi
#2}}
\providecommand{\BIBdecl}{\relax}
\BIBdecl

\bibitem{mann2020covid}
D.~M. Mann, J.~Chen, R.~Chunara, P.~A. Testa, and O.~Nov, ``Covid-19 transforms
  health care through telemedicine: evidence from the field,'' \emph{Journal of
  the American Medical Informatics Association}, vol.~27, no.~7, pp.
  1132--1135, 2020.

\bibitem{watson2016impact}
A.~R. Watson, ``Impact of the digital age on transforming healthcare,''
  \emph{Healthcare Information Management Systems: Cases, Strategies, and
  Solutions}, pp. 219--233, 2016.

\bibitem{erickson2017putting}
S.~M. Erickson, B.~Rockwern, M.~Koltov, R.~M. McLean, M.~Practice, and Q.~C.
  of~the American College~of Physicians*, ``Putting patients first by reducing
  administrative tasks in health care: a position paper of the american college
  of physicians,'' \emph{Annals of internal medicine}, vol. 166, no.~9, pp.
  659--661, 2017.

\bibitem{tutty2019complex}
M.~A. Tutty, L.~E. Carlasare, S.~Lloyd, and C.~A. Sinsky, ``The complex case of
  ehrs: examining the factors impacting the ehr user experience,''
  \emph{Journal of the American Medical Informatics Association}, vol.~26,
  no.~7, pp. 673--677, 2019.

\bibitem{abouzahra2014integrating}
M.~Abouzahra, K.~Sartipi, D.~Armstrong, and J.~Tan, ``Integrating data from
  ehrs to enhance clinical decision making: the inflammatory bowel disease
  case,'' in \emph{2014 IEEE 27th International Symposium on Computer-Based
  Medical Systems}.\hskip 1em plus 0.5em minus 0.4em\relax IEEE, 2014, pp.
  531--532.

\bibitem{ben2015improving}
O.~Ben-Assuli, D.~Sagi, M.~Leshno, A.~Ironi, and A.~Ziv, ``Improving diagnostic
  accuracy using ehr in emergency departments: A simulation-based study,''
  \emph{Journal of biomedical informatics}, vol.~55, pp. 31--40, 2015.

\bibitem{aldosari2017patients}
B.~Aldosari, ``Patients' safety in the era of emr/ehr automation,''
  \emph{Informatics in Medicine Unlocked}, vol.~9, pp. 230--233, 2017.

\bibitem{han2016effect}
J.~E. Han, M.~Rabinovich, P.~Abraham, P.~Satyanarayana, T.~V. Liao, T.~N.
  Udoji, G.~A. Cotsonis, E.~G. Honig, and G.~S. Martin, ``Effect of electronic
  health record implementation in critical care on survival and medication
  errors,'' \emph{The American journal of the medical sciences}, vol. 351,
  no.~6, pp. 576--581, 2016.

\bibitem{susnjak2023forecasting}
T.~Susnjak and P.~Maddigan, ``Forecasting patient demand at urgent care clinics
  using explainable machine learning,'' \emph{CAAI Transactions on Intelligence
  Technology}, vol.~8, no.~3, pp. 712--733, 2023.

\bibitem{paranjape2020short}
K.~Paranjape, M.~Schinkel, and P.~Nanayakkara, ``Short keynote paper:
  Mainstreaming personalized healthcare--transforming healthcare through new
  era of artificial intelligence,'' \emph{IEEE journal of biomedical and health
  informatics}, vol.~24, no.~7, pp. 1860--1863, 2020.

\bibitem{talpada2019analysis}
H.~Talpada, M.~N. Halgamuge, and N.~T.~Q. Vinh, ``An analysis on use of deep
  learning and lexical-semantic based sentiment analysis method on twitter data
  to understand the demographic trend of telemedicine,'' in \emph{2019 11th
  International Conference on Knowledge and Systems Engineering (KSE)}.\hskip
  1em plus 0.5em minus 0.4em\relax IEEE, 2019, pp. 1--9.

\bibitem{liu2023hierarchical}
T.~Liu, F.~Liu, Y.~Wan, R.~Hu, Y.~Zhu, and L.~Li, ``Hierarchical graph learning
  with convolutional network for brain disease prediction,'' \emph{Multimedia
  Tools and Applications}, pp. 1--19, 2023.

\bibitem{munasinghe2023supply}
U.~J. Munasinghe and M.~N. Halgamuge, ``Supply chain traceability and
  counterfeit detection of covid-19 vaccines using novel blockchain-based
  vacledger system,'' \emph{Expert Systems with Applications}, vol. 228, p.
  120293, 2023.

\bibitem{dagliati2021health}
A.~Dagliati, A.~Malovini, V.~Tibollo, and R.~Bellazzi, ``Health informatics and
  ehr to support clinical research in the covid-19 pandemic: an overview,''
  \emph{Briefings in bioinformatics}, vol.~22, no.~2, pp. 812--822, 2021.

\bibitem{osborne2020automated}
T.~F. Osborne, Z.~P. Veigulis, D.~M. Arreola, E.~R{\"o}{\"o}sli, and C.~M.
  Curtin, ``Automated ehr score to predict covid-19 outcomes at us department
  of veterans affairs,'' \emph{PLoS One}, vol.~15, no.~7, p. e0236554, 2020.

\bibitem{basil2022health}
N.~N. Basil, S.~Ambe, C.~Ekhator, E.~Fonkem, B.~N. Nduma, and C.~Ekhator,
  ``Health records database and inherent security concerns: A review of the
  literature,'' \emph{Cureus}, vol.~14, no.~10, 2022.

\bibitem{ganiga2020security}
R.~Ganiga, R.~M. Pai, and R.~K. Sinha, ``Security framework for cloud based
  electronic health record (ehr) system,'' \emph{International Journal of
  Electrical and Computer Engineering}, vol.~10, no.~1, p. 455, 2020.

\bibitem{mcintosh2019masquerade}
T.~McIntosh, J.~Jang-Jaccard, P.~Watters, and T.~Susnjak, ``Masquerade attacks
  against security software exclusion lists,'' \emph{Australian Journal of
  Intelligent Information Processing Systems}, vol.~16, no.~4, pp. 5--12, 2019.

\bibitem{mcintosh2021ransomware}
T.~McIntosh, A.~Kayes, Y.-P.~P. Chen, A.~Ng, and P.~Watters, ``Ransomware
  mitigation in the modern era: A comprehensive review, research challenges,
  and future directions,'' \emph{ACM Computing Surveys (CSUR)}, vol.~54, no.~9,
  pp. 1--36, 2021.

\bibitem{vidanapathirana2022rapid}
D.~Vidanapathirana, A.~Mohammad, and M.~N. Halgamuge, ``Rapid cyber-attack
  detection system with low probability of missed attack warnings,'' in
  \emph{2022 IEEE 17th Conference on Industrial Electronics and Applications
  (ICIEA)}.\hskip 1em plus 0.5em minus 0.4em\relax IEEE, 2022, pp. 1423--1429.

\bibitem{mcintosh2023applying}
T.~McIntosh, A.~Kayes, Y.-P.~P. Chen, A.~Ng, and P.~Watters, ``Applying staged
  event-driven access control to combat ransomware,'' \emph{Computers \&
  Security}, vol. 128, p. 103160, 2023.

\bibitem{mcintosh2022intercepting}
T.~McIntosh, ``Intercepting ransomware attacks with staged event-driven access
  control,'' Ph.D. dissertation, La Trobe, 2022.

\bibitem{fernandez2013security}
J.~L. Fern{\'a}ndez-Alem{\'a}n, I.~C. Se{\~n}or, P.~{\'A}.~O. Lozoya, and
  A.~Toval, ``Security and privacy in electronic health records: A systematic
  literature review,'' \emph{Journal of biomedical informatics}, vol.~46,
  no.~3, pp. 541--562, 2013.

\bibitem{rezaeibagha2015systematic}
F.~Rezaeibagha, K.~T. Win, and W.~Susilo, ``A systematic literature review on
  security and privacy of electronic health record systems: technical
  perspectives,'' \emph{Health Information Management Journal}, vol.~44, no.~3,
  pp. 23--38, 2015.

\bibitem{liu2015auditing}
W.~Liu, X.~Liu, J.~Liu, Q.~Wu, J.~Zhang, and Y.~Li, ``Auditing and revocation
  enabled role-based access control over outsourced private ehrs,'' in
  \emph{2015 IEEE 17th international conference on high performance computing
  and communications, 2015 IEEE 7th international symposium on cyberspace
  safety and security, and 2015 IEEE 12th international conference on embedded
  software and systems}.\hskip 1em plus 0.5em minus 0.4em\relax IEEE, 2015, pp.
  336--341.

\bibitem{abirami2019attribute}
G.~Abirami and R.~Venkataraman, ``Attribute based access control with trust
  calculation (abac-t) for decision policies of health care in pervasive
  environment,'' \emph{IJITEE}, vol.~8, 2019.

\bibitem{psarra2020securing}
E.~Psarra, I.~Patiniotakis, Y.~Verginadis, D.~Apostolou, and G.~Mentzas,
  ``Securing access to healthcare data with context-aware policies,'' in
  \emph{2020 11th International Conference on Information, Intelligence,
  Systems and Applications (IISA}.\hskip 1em plus 0.5em minus 0.4em\relax IEEE,
  2020, pp. 1--6.

\bibitem{kopanitsa2017integration}
G.~Kopanitsa, ``Integration of hospital information and clinical decision
  support systems to enable the reuse of electronic health record data,''
  \emph{Methods of information in medicine}, vol.~56, no.~4, pp. 238--247,
  2017.

\bibitem{adel2019unified}
E.~Adel, S.~El-Sappagh, S.~Barakat, and M.~Elmogy, ``A unified fuzzy ontology
  for distributed electronic health record semantic interoperability,'' in
  \emph{U-Healthcare Monitoring Systems}.\hskip 1em plus 0.5em minus
  0.4em\relax Elsevier, 2019, pp. 353--395.

\bibitem{fragidis2018implementation}
L.~L. Fragidis and P.~D. Chatzoglou, ``Implementation of a nationwide
  electronic health record (ehr): The international experience in 13
  countries,'' \emph{International journal of health care quality assurance},
  vol.~31, no.~2, pp. 116--130, 2018.

\bibitem{mcintosh2023harnessing}
T.~McIntosh, T.~Liu, T.~Susnjak, H.~Alavizadeh, A.~Ng, R.~Nowrozy, and
  P.~Watters, ``Harnessing gpt-4 for generation of cybersecurity grc policies:
  A focus on ransomware attack mitigation,'' \emph{Computers \& Security}, vol.
  134, p. 103424, 2023.

\bibitem{ont_security}
E.~Ghazizadeh, E.~Bagheri, and P.~M. Singh, ``Security ontology for electronic
  health records,'' \emph{Journal of biomedical informatics}, vol.~53, pp.
  196--207, 2015.

\bibitem{ca_abac_integration}
E.~Vergara and J.~Lopez, ``Context-aware attribute-based access control,'' in
  \emph{International Conference on Information Security and Cryptology}.\hskip
  1em plus 0.5em minus 0.4em\relax Springer, 2013, pp. 165--180.

\bibitem{ont_integration}
J.~He, X.~Chen, J.~Zhang, and J.~Yu, ``An ontology-driven approach for securing
  electronic health records,'' \emph{BMC medical informatics and decision
  making}, vol.~13, no.~1, p.~12, 2013.

\bibitem{ont_ca_abac_integration}
H.~Liu, S.~Yu, and X.~Yang, ``Ontology-driven context-aware attribute-based
  access control model for healthcare applications,'' \emph{Journal of medical
  systems}, vol.~42, no.~12, p. 249, 2018.

\bibitem{ntalasha2019adaptive}
D.~Ntalasha, R.~Li, and Y.~Wang, ``Adaptive context-aware design using context
  state information for the internet of things paradigm,'' \emph{Journal of
  Mobile Multimedia}, pp. 289--320, 2019.

\bibitem{sicuranza2013access}
M.~Sicuranza and A.~Esposito, ``An access control model for easy management of
  patient privacy in ehr systems,'' in \emph{8th International Conference for
  Internet Technology and Secured Transactions (ICITST-2013)}.\hskip 1em plus
  0.5em minus 0.4em\relax IEEE, 2013, pp. 463--470.

\bibitem{de2018health}
M.~A. de~Carvalho~Junior, P.~Bandiera-Paiva \emph{et~al.}, ``Health information
  system role-based access control current security trends and challenges,''
  \emph{Journal of healthcare engineering}, vol. 2018, 2018.

\bibitem{zhang2014role}
R.~Zhang, L.~Liu, and R.~Xue, ``Role-based and time-bound access and management
  of ehr data,'' \emph{Security and communication Networks}, vol.~7, no.~6, pp.
  994--1015, 2014.

\bibitem{esposito2013patient}
A.~Esposito, M.~Sicuranza, and M.~Ciampi, ``A patient centric approach for
  modeling access control in ehr systems,'' in \emph{Algorithms and
  Architectures for Parallel Processing: 13th International Conference, ICA3PP
  2013, Vietri sul Mare, Italy, December 18-20, 2013, Proceedings, Part II
  13}.\hskip 1em plus 0.5em minus 0.4em\relax Springer, 2013, pp. 225--232.

\bibitem{santos2013secure}
C.~Santos-Pereira, A.~B. Augusto, R.~Cruz-Correia, and M.~E. Correia, ``A
  secure rbac mobile agent access control model for healthcare institutions,''
  in \emph{Proceedings of the 26th IEEE international symposium on
  computer-based medical systems}.\hskip 1em plus 0.5em minus 0.4em\relax IEEE,
  2013, pp. 349--354.

\bibitem{sicuranza2015view}
M.~Sicuranza, A.~Esposito, and M.~Ciampi, ``A view-based acces control model
  for ehr systems,'' in \emph{Intelligent Distributed Computing VIII}.\hskip
  1em plus 0.5em minus 0.4em\relax Springer, 2015, pp. 443--452.

\bibitem{liu2017auditing}
W.~Liu, X.~Liu, J.~Liu, and Q.~Wu, ``Auditing revocable privacy-preserving
  access control for ehrs in clouds,'' \emph{The Computer Journal}, vol.~60,
  no.~12, pp. 1871--1888, 2017.

\bibitem{chen2012risk}
L.~Chen, M.~J. Kollingbaum, T.~J. Norman, and P.~Edwards, ``Risk-aware access
  control for electronic health records,'' in \emph{Proceedings of the Third
  Annual Digital Economy All Hands Conference, Aberdeen}, 2012.

\bibitem{abouelmehdi2018big}
K.~Abouelmehdi, A.~Beni-Hessane, and H.~Khaloufi, ``Big healthcare data:
  preserving security and privacy,'' \emph{Journal of big data}, vol.~5, no.~1,
  pp. 1--18, 2018.

\bibitem{zarezadeh2020attribute}
M.~Zarezadeh, M.~A. Taluki, and M.~Siavashi, ``Attribute-based access control
  for cloud-based electronic health record (ehr) systems.'' \emph{ISeCure},
  vol.~12, no.~2, 2020.

\bibitem{alshiky2017attribute}
A.~M. Alshiky, S.~M. Buhari, and A.~Barnawi, ``Attribute-based access control
  (abac) for ehr in fog computing environment,'' \emph{International Journal on
  Cloud Computing: Services and Architecture (IJCCSA)}, vol.~7, no.~1, pp.
  27--34, 2017.

\bibitem{sahavechaphan2012efficient}
N.~Sahavechaphan, U.~Suriya, N.~Harnsamut, J.~Phengsuwan, K.~Aroonrua
  \emph{et~al.}, ``An efficient technique for aspect-based ehr access policy
  administration on abac,'' in \emph{2011 Ninth International Conference on ICT
  and Knowledge Engineering}.\hskip 1em plus 0.5em minus 0.4em\relax IEEE,
  2012, pp. 27--33.

\bibitem{joshi2018attribute}
M.~Joshi, K.~Joshi, and T.~Finin, ``Attribute based encryption for secure
  access to cloud based ehr systems,'' in \emph{2018 IEEE 11th International
  Conference on Cloud Computing (CLOUD)}.\hskip 1em plus 0.5em minus
  0.4em\relax IEEE, 2018, pp. 932--935.

\bibitem{guo2019access}
H.~Guo, W.~Li, M.~Nejad, and C.-C. Shen, ``Access control for electronic health
  records with hybrid blockchain-edge architecture,'' in \emph{2019 IEEE
  International Conference on Blockchain (Blockchain)}.\hskip 1em plus 0.5em
  minus 0.4em\relax IEEE, 2019, pp. 44--51.

\bibitem{walid2023semantically}
R.~Walid, K.~P. Joshi, and S.~G. Choi, ``Semantically rich differential access
  to secure cloud ehr,'' in \emph{2023 IEEE 9th Intl Conference on Big Data
  Security on Cloud (BigDataSecurity), IEEE Intl Conference on High Performance
  and Smart Computing,(HPSC) and IEEE Intl Conference on Intelligent Data and
  Security (IDS)}.\hskip 1em plus 0.5em minus 0.4em\relax IEEE, 2023, pp. 1--9.

\bibitem{seol2018privacy}
K.~Seol, Y.-G. Kim, E.~Lee, Y.-D. Seo, and D.-K. Baik, ``Privacy-preserving
  attribute-based access control model for xml-based electronic health record
  system,'' \emph{IEEE Access}, vol.~6, pp. 9114--9128, 2018.

\bibitem{patra2022controlling}
L.~Patra, U.~P. Rao, P.~Choksi, and A.~Chaurasia, ``Controlling access to
  ehealth data using request denial cache in xacml reference architecture for
  abac,'' in \emph{2022 IEEE 3rd Global Conference for Advancement in
  Technology (GCAT)}.\hskip 1em plus 0.5em minus 0.4em\relax IEEE, 2022, pp.
  1--8.

\bibitem{arfaoui2019context}
A.~Arfaoui, S.~Cherkaoui, A.~Kribeche, S.~M. Senouci, and M.~Hamdi,
  ``Context-aware adaptive authentication and authorization in internet of
  things,'' in \emph{ICC 2019-2019 IEEE International Conference on
  Communications (ICC)}.\hskip 1em plus 0.5em minus 0.4em\relax IEEE, 2019, pp.
  1--6.

\bibitem{el2020survey}
R.~El~Sibai, N.~Gemayel, J.~Bou~Abdo, and J.~Demerjian, ``A survey on access
  control mechanisms for cloud computing,'' \emph{Transactions on Emerging
  Telecommunications Technologies}, vol.~31, no.~2, p. e3720, 2020.

\bibitem{chen2011novel}
L.~Chen and D.~B. Hoang, ``Novel data protection model in healthcare cloud,''
  in \emph{2011 IEEE International Conference on High Performance Computing and
  Communications}.\hskip 1em plus 0.5em minus 0.4em\relax IEEE, 2011, pp.
  550--555.

\bibitem{padmapriya2021preserving}
S.~Padmapriya, R.~Shankar, R.~Thiagarajan, S.~Arun, B.~Liya, and
  B.~Gunasundari, ``Preserving privacy scheme using data-caac mechanism in
  e-health based on hybrid edge computing,'' in \emph{2021 3rd International
  Conference on Advances in Computing, Communication Control and Networking
  (ICAC3N)}.\hskip 1em plus 0.5em minus 0.4em\relax IEEE, 2021, pp. 1394--1399.

\bibitem{kayes2015ontcaac}
A.~Kayes, J.~Han, and A.~Colman, ``Ontcaac: an ontology-based approach to
  context-aware access control for software services,'' \emph{The Computer
  Journal}, vol.~58, no.~11, pp. 3000--3034, 2015.

\bibitem{yarmand2008behavior}
M.~H. Yarmand, K.~Sartipi, and D.~G. Down, ``Behavior-based access control for
  distributed healthcare environment,'' in \emph{2008 21st IEEE International
  Symposium on Computer-Based Medical Systems}.\hskip 1em plus 0.5em minus
  0.4em\relax IEEE, 2008, pp. 126--131.

\bibitem{yarmand2013behavior}
------, ``Behavior-based access control for distributed healthcare systems,''
  \emph{Journal of Computer Security}, vol.~21, no.~1, pp. 1--39, 2013.

\bibitem{ke2021privacy}
C.~Ke, J.~Wu, F.~Xiao, Z.~Huang, and Y.~Meng, ``A privacy risk assessment
  scheme for fog nodes in access control system,'' \emph{IEEE Transactions on
  Reliability}, vol.~71, no.~4, pp. 1513--1526, 2021.

\bibitem{sicuranza2014semantic}
M.~Sicuranza and M.~Ciampi, ``A semantic access control for easy management of
  the privacy for ehr systems,'' in \emph{2014 Ninth International Conference
  on P2P, Parallel, Grid, Cloud and Internet Computing}.\hskip 1em plus 0.5em
  minus 0.4em\relax IEEE, 2014, pp. 400--405.

\bibitem{calvillo2014standardized}
J.~Calvillo-Arbizu, I.~Rom{\'a}n-Mart{\'\i}nez, and L.~M. Roa-Romero,
  ``Standardized access control mechanisms for protecting iso 13606-based
  electronic health record systems,'' in \emph{IEEE-EMBS International
  Conference on Biomedical and Health Informatics (BHI)}.\hskip 1em plus 0.5em
  minus 0.4em\relax IEEE, 2014, pp. 539--542.

\bibitem{dixit2019multi}
S.~Dixit, K.~P. Joshi, and S.~G. Choi, ``Multi authority access control in a
  cloud ehr system with ma-abe,'' in \emph{2019 IEEE international conference
  on edge computing (EDGE)}.\hskip 1em plus 0.5em minus 0.4em\relax IEEE, 2019,
  pp. 107--109.

\bibitem{walid2020cloud}
R.~Walid, K.~P. Joshi, S.~G. Choi, and D.-y. Kim, ``Cloud-based encrypted ehr
  system with semantically rich access control and searchable encryption,'' in
  \emph{2020 IEEE International Conference on Big Data (Big Data)}.\hskip 1em
  plus 0.5em minus 0.4em\relax IEEE, 2020, pp. 4075--4082.

\bibitem{peleg2008situation}
M.~Peleg, D.~Beimel, D.~Dori, and Y.~Denekamp, ``Situation-based access
  control: Privacy management via modeling of patient data access scenarios,''
  \emph{Journal of Biomedical Informatics}, vol.~41, no.~6, pp. 1028--1040,
  2008.

\bibitem{beimel2009reasoning}
D.~Beimel, M.~Peleg, and T.~Redmond, ``Reasoning about access-control
  situations with owl,'' in \emph{The 11th Intl Prot{\'e}g{\'e} Conference,
  Amsterdam, Netherlands}, 2009.

\bibitem{dong2015coc}
X.~Dong, R.~Samavi, and T.~Topaloglou, ``Coc: An ontology for capturing
  semantics of circle of care,'' \emph{Procedia Computer Science}, vol.~63, pp.
  589--594, 2015.

\bibitem{nowrozy2023towards}
R.~Nowrozy, A.~Khandakar, W.~Hua, and T.~Mcintosh, ``Towards a universal
  privacy model for electronic health record systems: An ontology and machine
  learning approach,'' \emph{Informatics}, vol.~10, no.~3, 2023.

\bibitem{mesko2023imperative}
B.~Mesk{\'o} and E.~J. Topol, ``The imperative for regulatory oversight of
  large language models (or generative ai) in healthcare,'' \emph{NPJ digital
  medicine}, vol.~6, no.~1, p. 120, 2023.

\bibitem{gupta2023chatgpt}
M.~Gupta, C.~Akiri, K.~Aryal, E.~Parker, and L.~Praharaj, ``From chatgpt to
  threatgpt: Impact of generative ai in cybersecurity and privacy,'' \emph{IEEE
  Access}, 2023.

\bibitem{tan2023generative}
T.~F. Tan, A.~J. Thirunavukarasu, J.~P. Campbell, P.~A. Keane, L.~R. Pasquale,
  M.~D. Abramoff, J.~Kalpathy-Cramer, F.~Lum, J.~E. Kim, S.~L. Baxter
  \emph{et~al.}, ``Generative artificial intelligence through chatgpt and other
  large language models in ophthalmology: Clinical applications and
  challenges,'' \emph{Ophthalmology Science}, vol.~3, no.~4, p. 100394, 2023.

\bibitem{molloy2012generative}
I.~Molloy, Y.~Park, and S.~Chari, ``Generative models for access control
  policies: applications to role mining over logs with attribution,'' in
  \emph{Proceedings of the 17th ACM symposium on Access Control Models and
  Technologies}, 2012, pp. 45--56.

\bibitem{solaiman2023gradient}
I.~Solaiman, ``The gradient of generative ai release: Methods and
  considerations,'' in \emph{Proceedings of the 2023 ACM Conference on
  Fairness, Accountability, and Transparency}, 2023, pp. 111--122.

\bibitem{mccoy2022believing}
L.~G. McCoy, C.~T. Brenna, S.~S. Chen, K.~Vold, and S.~Das, ``Believing in
  black boxes: machine learning for healthcare does not need explainability to
  be evidence-based,'' \emph{Journal of clinical epidemiology}, vol. 142, pp.
  252--257, 2022.

\bibitem{felzmann2020towards}
H.~Felzmann, E.~Fosch-Villaronga, C.~Lutz, and A.~Tam{\`o}-Larrieux, ``Towards
  transparency by design for artificial intelligence,'' \emph{Science and
  Engineering Ethics}, vol.~26, no.~6, pp. 3333--3361, 2020.

\bibitem{khattak2023ethical}
W.~A. Khattak and F.~Rabbi, ``Ethical considerations and challenges in the
  deployment of natural language processing systems in healthcare,''
  \emph{International Journal of Applied Health Care Analytics}, vol.~8, no.~5,
  pp. 17--36, 2023.

\bibitem{sison2023chatgpt}
A.~J.~G. Sison, M.~T. Daza, R.~Gozalo-Brizuela, and E.~C. Garrido-Merch{\'a}n,
  ``Chatgpt: More than a weapon of mass deception, ethical challenges and
  responses from the human-centered artificial intelligence (hcai)
  perspective,'' \emph{arXiv preprint arXiv:2304.11215}, 2023.

\bibitem{zerkouk2014behavior}
M.~Zerkouk, P.~Cavalcante, A.~Mhamed, J.~Boudy, and B.~Messabih, ``Behavior and
  capability based access control model for personalized telehealthcare
  assistance,'' \emph{Mobile Networks and Applications}, vol.~19, pp. 392--403,
  2014.

\bibitem{mustapha4394368systematic}
A.~M. Mustapha, T.~E. Abioye, O.~Oyedele, F.~M. Okikiola, and C.~Y. Alonge, ``A
  systematic literature review of ontology-based techniques in medical
  diagnosis,'' \emph{Available at SSRN 4394368}.

\bibitem{sharma2016ontology}
K.~Sharma, S.~Gupta, R.~Kaur, and M.~Kumar, ``Ontology driven electronic health
  record,'' in \emph{2016 International Conference on Computing, Communication
  and Automation (ICCCA)}.\hskip 1em plus 0.5em minus 0.4em\relax IEEE, 2016,
  pp. 940--944.

\bibitem{tall2023framework}
A.~M. Tall and C.~C. Zou, ``A framework for attribute-based access control in
  processing big data with multiple sensitivities,'' \emph{Applied Sciences},
  vol.~13, no.~2, p. 1183, 2023.

\bibitem{dhillon2023extended}
P.~Dhillon and M.~Singh, ``An extended ontology model for trust evaluation
  using advanced hybrid ontology,'' \emph{Journal of Information Science}, p.
  01655515221128424, 2023.

\bibitem{wahlberg2017legal}
L.~Wahlberg, ``Legal ontology, scientific expertise and the factual world,''
  \emph{Journal of Social Ontology}, vol.~3, no.~1, pp. 49--65, 2017.

\bibitem{kiong2011health}
Y.~C. Kiong, S.~Palaniappan, and N.~A. Yahaya, ``Health ontology system,'' in
  \emph{2011 7th International Conference on Information Technology in
  Asia}.\hskip 1em plus 0.5em minus 0.4em\relax IEEE, 2011, pp. 1--4.

\bibitem{helms2011evaluating}
E.~Helms and L.~Williams, ``Evaluating access control of open source electronic
  health record systems,'' in \emph{Proceedings of the 3rd workshop on software
  engineering in health care}, 2011, pp. 63--70.

\bibitem{yang2022multiple}
Y.~Yang, R.-h. Shi, K.~Li, Z.~Wu, and S.~Wang, ``Multiple access control scheme
  for ehrs combining edge computing with smart contracts,'' \emph{Future
  Generation Computer Systems}, vol. 129, pp. 453--463, 2022.

\bibitem{kruse2017}
C.~S. Kruse, B.~Frederick, T.~Jacobson, and D.~K. Monticone, ``Cybersecurity in
  healthcare: A systematic review of modern threats and trends,''
  \emph{Technology and Health Care}, vol.~25, no.~1, pp. 1--10, 2017.

\bibitem{mcintosh2023google}
T.~R. McIntosh, T.~Susnjak, T.~Liu, P.~Watters, and M.~N. Halgamuge, ``From
  google gemini to openai q*(q-star): A survey of reshaping the generative
  artificial intelligence (ai) research landscape,'' \emph{arXiv preprint
  arXiv:2312.10868}, 2023.

\bibitem{zhang2017}
Y.~Zhang and M.~Yang, ``Intelligent cloud storage usage for electronic health
  record system,'' \emph{Journal of Medical Systems}, vol.~41, no.~3, p.~44,
  2017.

\bibitem{ghanbari2018}
S.~Ghanbari and M.~A. Azgomi, ``A taxonomy and survey of cloud resource
  orchestration techniques,'' \emph{ACM Computing Surveys (CSUR)}, vol.~51,
  no.~3, pp. 1--34, 2018.

\bibitem{papakonstantinou2016}
V.~Papakonstantinou, M.~Poulymenopoulou, F.~Malamateniou, and
  G.~Vassilacopoulos, ``Access control for cloud-based emergency medical data
  management systems,'' \emph{Health Informatics Journal}, vol.~22, no.~4, pp.
  812--824, 2016.

\bibitem{chintagunta2021medically}
B.~Chintagunta, N.~Katariya, X.~Amatriain, and A.~Kannan, ``Medically aware
  gpt-3 as a data generator for medical dialogue summarization,'' in
  \emph{Machine Learning for Healthcare Conference}.\hskip 1em plus 0.5em minus
  0.4em\relax PMLR, 2021, pp. 354--372.

\bibitem{korngiebel2021considering}
D.~M. Korngiebel and S.~D. Mooney, ``Considering the possibilities and pitfalls
  of generative pre-trained transformer 3 (gpt-3) in healthcare delivery,''
  \emph{NPJ Digital Medicine}, vol.~4, no.~1, p.~93, 2021.

\bibitem{lee2021application}
D.~Lee and S.~N. Yoon, ``Application of artificial intelligence-based
  technologies in the healthcare industry: Opportunities and challenges,''
  \emph{International Journal of Environmental Research and Public Health},
  vol.~18, no.~1, p. 271, 2021.

\bibitem{moghaddam2020future}
Y.~Moghaddam, H.~Yurko, H.~Demirkan, N.~Tymann, and A.~Rayes, \emph{The future
  of work: how artificial intelligence can augment human capabilities}.\hskip
  1em plus 0.5em minus 0.4em\relax business expert press, 2020.

\bibitem{yeung2018study}
K.~Yeung, ``A study of the implications of advanced digital technologies
  (including ai systems) for the concept of responsibility within a human
  rights framework,'' \emph{MSI-AUT (2018)}, vol.~5, 2018.

\bibitem{brandao2013social}
C.~Brand{\~a}o, G.~Rego, I.~Duarte, and R.~Nunes, ``Social responsibility: a
  new paradigm of hospital governance?'' \emph{Health Care Analysis}, vol.~21,
  pp. 390--402, 2013.

\bibitem{aydin2021}
M.~N. Aydin and A.~Ali, ``Ethical and security challenges in electronic health
  records: A review,'' \emph{Journal of medical systems}, vol.~45, no.~8,
  p.~90, 2021.

\bibitem{shadish2021}
W.~R. Shadish, T.~D. Cook, and D.~T. Campbell, ``Experimental and
  quasi-experimental designs for generalized causal inference,'' 2021.

\bibitem{yao2021}
Y.~Yao, H.~Wang, and Y.~Li, ``A novel security and privacy-preserving scheme
  for electronic health record systems,'' \emph{IEEE Access}, vol.~9, pp.
  70\,524--70\,536, 2021.

\bibitem{gupta2020}
S.~Gupta, A.~Basheeruddin, and P.~Kumar, ``A systematic review on information
  security risks and threats in healthcare information systems,''
  \emph{Computers in Biology and Medicine}, vol. 118, p. 103641, 2020.

\bibitem{ando2022}
H.~Ando, M.~Ohkubo, and K.~Ikeda, ``Information security governance and
  management in healthcare: A systematic literature review,''
  \emph{International Journal of Medical Informatics}, vol. 157, p. 104608,
  2022.

\bibitem{wang2020}
D.~Wang and A.~Bakhai, ``Randomized controlled trials: design, conduct, and
  analysis,'' \emph{The Lancet}, vol. 395, no. 10223, pp. 1316--1325, 2020.

\bibitem{pereira2021}
V.~Pereira and F.~Santos, ``An ontology-based approach for managing security in
  electronic health records,'' \emph{Journal of biomedical informatics}, vol.
  118, p. 103792, 2021.

\bibitem{yuan2021}
X.~Yuan and J.~Huang, ``Securing electronic health records using blockchain
  technology: A systematic review,'' \emph{Journal of medical systems},
  vol.~45, no.~5, p.~49, 2021.

\bibitem{zhang2020privacy}
Y.~Zhang, B.~Xie, M.~Zhang, Q.~Cui, and L.~Xie, ``Privacy preservation in
  electronic health records: A survey,'' \emph{Journal of medical systems},
  vol.~44, no.~4, pp. 1--13, 2020.

\bibitem{mcintosh2023culturally}
T.~R. McIntosh, T.~Liu, T.~Susnjak, P.~Watters, A.~Ng, and M.~N. Halgamuge, ``A
  culturally sensitive test to evaluate nuanced gpt hallucination,'' \emph{IEEE
  Transactions on Artificial Intelligence}, vol.~1, no.~01, pp. 1--13, 2023.

\bibitem{abd2021access}
M.~Abd and Y.~Ma, ``Access control and privacy protection in healthcare
  information systems: A systematic literature review,'' \emph{Journal of
  Healthcare Engineering}, vol. 2021, 2021.

\bibitem{1874376:27758905}
S.~Kumar, K.~S, J.~Hanumanthappa, S.~P.~S. Prakash, and K.~Krinkin,
  ``{Relationship-Based AES Security Model for Social Internet of Things},'' in
  \emph{{Intelligent Systems and Applications: Select Proceedings of ICISA
  2022}}.\hskip 1em plus 0.5em minus 0.4em\relax Springer Nature, 2023, pp.
  143--151.

\bibitem{1874376:27758933}
P.~Ruotsalainen and B.~Blobel, \emph{{Health information systems in the digital
  health ecosystem-problems and solutions for ethics, trust and privacy}},
  vol.~17, pp. 3006--3006, 2020.

\bibitem{1874376:27758956}
G.~G{\"{a}}bler, D.~Lycett, and W.~Gall, ``{Integrating a New Dietetic Care
  Process in a Health Information System: A System and Process Analysis and
  Assessment},'' \emph{{International Journal of Environmental Research and
  Public Health}}, vol.~19, no.~5, pp. 2491--2491, 2022.

\bibitem{huang2019ontology}
C.-C. Huang and C.-L. Tsai, ``Ontology-based access control for electronic
  health records: A survey,'' \emph{Journal of medical systems}, vol.~43,
  no.~9, p. 297, 2019.

\bibitem{dhanapal2021secure}
M.~Dhanapal and Y.~G, ``Secure medical record management using blockchain
  technology in cloud environment,'' \emph{Journal of Medical Systems},
  vol.~45, no.~2, p.~13, 2021.

\bibitem{DPHealthcare2021}
A.~Name, ``Title of the differential privacy in healthcare paper,''
  \emph{Journal Name}, 2021.

\bibitem{ethical_decision_paper}
E.~Author and F.~Author, ``Ethical considerations in ehr security,''
  \emph{International Journal of Health Ethics}, vol.~15, no.~1, pp. 45--60,
  2019.

\bibitem{coorevits2013electronic}
P.~Coorevits, M.~Sundgren, G.~O. Klein, A.~Bahr, B.~Claerhout, C.~Daniel,
  M.~Dugas, D.~Dupont, A.~Schmidt, P.~Singleton \emph{et~al.}, ``Electronic
  health records: new opportunities for clinical research,'' \emph{Journal of
  internal medicine}, vol. 274, no.~6, pp. 547--560, 2013.

\bibitem{carter2022electronic}
A.~B. Carter, L.~V. Abruzzo, J.~W. Hirschhorn, D.~Jones, D.~C. Jordan,
  M.~Nassiri, S.~Ogino, N.~R. Patel, C.~G. Suciu, R.~L. Temple-Smolkin
  \emph{et~al.}, ``Electronic health records and genomics: perspectives from
  the association for molecular pathology electronic health record (ehr)
  interoperability for clinical genomics data working group,'' \emph{The
  Journal of Molecular Diagnostics}, vol.~24, no.~1, pp. 1--17, 2022.

\bibitem{wang2018towards}
R.~Wang, Y.~Guo, Y.~Li, Z.~Qin, Y.~Huang, and Z.~Li, ``Towards personalized and
  privacy-preserving ehealth systems via semi-supervised learning,''
  \emph{Journal of medical systems}, vol.~42, no.~7, p. 129, 2018.

\bibitem{bhuyan2021role}
M.~Bhuyan, A.~Pal, and R.~Barik, ``Role based access control for secure data
  sharing in cloud using cloud computing,'' in \emph{Proceedings of the 11th
  International Conference on Computing, Communication and Networking
  Technologies (ICCCNT)}, 2021, pp. 1--5.

\end{thebibliography}

\begin{IEEEbiography}[{\includegraphics[width=1in,height=1.25in,clip,keepaspectratio]{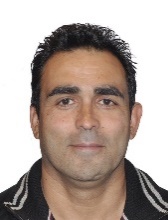}}]{Raza Nowrozy} is an accomplished cybersecurity professional with extensive industry experience, holding a B.Sc in Technology from the University of South Australia and a Master of Cybersecurity from La Trobe University. He is currently pursuing a PhD in Cybersecurity at Victoria University, focusing on privacy and security within Health Information, with an emphasis on My Health Record (MHR). As a certified cyber trainer with several professional certifications, Raza's unique combination of academic qualifications, professional credentials, and practical experience sets him apart as a highly skilled cybersecurity authority. He is steadfastly committed to research and training initiatives, demonstrating his dedication to promoting best practices and fostering the development of the next generation of cybersecurity professionals.
\end{IEEEbiography}

\begin{IEEEbiography}[{\includegraphics[width=1in,height=1.25in,clip,keepaspectratio]{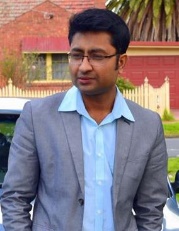}}]{Dr Khandakar Ahmed} who completed his undergraduate degree at his current place of employment, Victoria University, began his career as a full-stack developer before transitioning to academia. He taught at multiple universities in Europe and Australia before working as a post-doctoral research fellow at RMIT in 2015~2016 and eventually joining Victoria University in 2017 as a Senior Lecturer in IT at the College of Engineering and Science. Khandakar has participated in numerous research projects, collaborating with industries and various levels of government to use intelligent technologies to solve contemporary social issues, with a focus on the Internet of Things, smart cities, machine learning, cybersecurity, and biomedical informatics. He has received substantial funding from industry partners over the past five years. 
\end{IEEEbiography}

\begin{IEEEbiography}[{\includegraphics[width=1in,height=1.25in,clip,keepaspectratio]{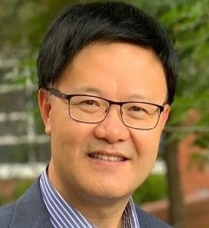}}]{Professor Hua Wang} (Senior Member, IEEE) received his Ph.D. degree from the University of Southern Queensland, Toowoomba, Qld., Australia in 2004. He is now a full time Professor at Victoria University. He has expertise in electronic commerce, business process modeling, and enterprise architecture. As a Chief Investigator, six Australian Research Council (ARC) Discovery grants have been awarded since 2006, and 350 peer-reviewed scholar papers have been published. 16 Ph.D. students have already graduated under his principal supervision.
\end{IEEEbiography}
\vfill

\clearpage

\clearpage
\begin{appendices}
	\label{subsec:appendices}
	\section*{Appendix 1 - Questions for Evaluating Reasoning on the Raw Facts}
	\label{subsec:Appendix1}
	
	\subsection{\textbf{Category 1 - Allied Health}}
	\label{sec:Category 1}
	\begin{enumerate}
		
		\item \textbf{Dietitian:} A patient with recent weight loss due to Crohn's disease is referred to a dietitian. Can the dietitian, with the patient's consent, access the patient's My Health Record to understand the severity and duration of the disease to create a tailored nutritional plan, while ensuring the privacy and confidentiality of the patient's personal information?
		
		\item \textbf{Speech Therapist:} A child with delayed speech is referred to a speech therapist. Can the speech therapist, with the appropriate permissions, access the child's My Health Record to review any potential medical causes for the delay and tailor the therapy accordingly, while respecting the child's privacy?
		
		\item \textbf{Occupational Therapist:} A patient who has recently undergone a hip replacement surgery is assigned to an occupational therapist. Can the occupational therapist, with the necessary permissions, access the patient's My Health Record to understand the specifics of the surgery and design a personalized rehabilitation program, while maintaining the confidentiality of the patient's health information?
		
		\item \textbf{Physical Therapist:} A patient recovering from a severe sports injury is referred to a physical therapist. Can the physical therapist, with the patient's consent, access the patient's My Health Record to understand the full extent of the injury and create a recovery and strength-building plan, while ensuring the privacy of the patient's personal information?
		
		\item \textbf{Respiratory Therapist:} A patient with chronic obstructive pulmonary disease (COPD) is admitted to the hospital with breathing difficulty. Can the respiratory therapist, with the necessary permissions, access the patient's My Health Record to get detailed information about the patient's COPD history and provide the most effective respiratory support, while respecting the patient's privacy?
		
		\item \textbf{Audiologist:} A patient complaining of hearing loss is referred to an audiologist. Can the audiologist, with the patient's consent, access the patient's My Health Record to understand any medical issues that could be contributing to the hearing loss and plan the course of evaluation and treatment, while maintaining the confidentiality of the patient's health information?
		
		\item \textbf{Podiatrist:} A patient with long-standing diabetes presents with a foot ulcer. Can the podiatrist, with the necessary permissions, access the patient's My Health Record to review the history of their diabetes management and design a comprehensive treatment plan, while ensuring the privacy of the patient's personal information?
		
		\item \textbf{Radiation Therapist:} A patient diagnosed with prostate cancer is assigned a radiation therapist. Can the radiation therapist, with the patient's consent, access the patient's My Health Record to understand the specifics of the cancer and devise an optimal radiation therapy plan, while respecting the patient's privacy?
		
		\item \textbf{Orthotist:} A patient suffering from foot deformities due to rheumatoid arthritis is referred to an orthotist. Can the orthotist, with the necessary permissions, access the patient's My Health Record to assess the progression of the disease and design appropriate orthotic supports, while maintaining the confidentiality of the patient's health information?
		
		\item \textbf{Psychiatric Technician:} A patient with a history of psychiatric disorders is admitted to the hospital after an acute episode. Can the psychiatric technician, with the patient's consent, access the patient's My Health Record to understand their past treatments and medications, ensuring a safe and efficient care plan, while ensuring the privacy of the patient's personal information?
	\end{enumerate}
	
	\subsection{\textbf{Category 2 - Consultants}}
	\label{sec:Category 2}
	\begin{enumerate}
		
		\item \textbf{Endocrinologist:} A patient with uncontrolled diabetes is referred to an endocrinologist for further evaluation. Can the endocrinologist, with the patient's consent, access the patient's My Health Record to gain insight into the patient's historical glucose control and current treatment regimen, while ensuring the privacy and confidentiality of the patient's personal information?
		
		\item \textbf{Neurologist:} A patient who recently suffered a stroke is referred to a neurologist at a different hospital for follow-up care. Can the neurologist, with the appropriate permissions, access the patient's My Health Record to understand the severity of the stroke and plan further management, while respecting the patient's privacy?
		
		\item \textbf{Physical Therapist:} A patient recovering from a hip replacement surgery has been assigned a physical therapist for rehabilitation. Can the physical therapist, who works in a different clinic and with the necessary permissions, access the patient's My Health Record to comprehend the specifics of the surgery and design a tailored rehabilitation program, while maintaining the confidentiality of the patient's health information?
		
		\item \textbf{Cardiologist:} A patient's primary care doctor refers the patient to a cardiologist for an irregular heartbeat. Can the cardiologist, with the patient's consent, access the patient's My Health Record to review the patient's medical history and the details of any previous cardiac evaluations, while ensuring the privacy of the patient's personal information?
		
		\item \textbf{Nephrologist:} A patient with declining kidney function is referred to a nephrologist. Can the nephrologist, with the necessary permissions, access the patient's My Health Record to understand the trend of the patient's kidney function and devise a management plan, while respecting the patient's privacy?
		
		\item \textbf{Psychiatrist:} A patient with depression is referred by their primary care doctor to a psychiatrist for specialized care. Can the psychiatrist, with the patient's consent, access the patient's My Health Record to review past treatments and to formulate a comprehensive mental health management plan, while maintaining the confidentiality of the patient's health information?
		
		\item \textbf{Dermatologist:} A patient with a suspicious skin lesion is referred to a dermatologist for further evaluation. Can the dermatologist, with the necessary permissions, access the patient's My Health Record to check for the patient's history of sun exposure and previous skin conditions, while ensuring the privacy of the patient's personal information?
		
		\item \textbf{Rheumatologist:} A patient showing symptoms of an autoimmune disease is referred to a rheumatologist. Can the rheumatologist, with the patient's consent, access the patient's My Health Record to understand the progression of symptoms and guide their diagnosis and treatment approach, while respecting the patient's privacy?
		
		\item \textbf{Gastroenterologist:} A patient with chronic stomach pain is referred to a gastroenterologist. Can the gastroenterologist, with the necessary permissions, access the patient's My Health Record to gain insight into the patient's previous evaluations and guide further investigations, while maintaining the confidentiality of the patient's health information?
		
		\item \textbf{Pulmonologist:} A patient with chronic obstructive pulmonary disease (COPD) experiencing worsening symptoms is referred to a pulmonologist. Can the pulmonologist, with the patient's consent, access the patient's My Health Record to review the patient's history of COPD and adjust the treatment plan accordingly, while ensuring the privacy of the patient's personal information?
	\end{enumerate}
	
	\subsection{\textbf{Category 3 - Direct Care}}
	\label{sec:Category 3}
	\begin{enumerate}
		
		\item \textbf{General Practitioner:} A patient visits their general practitioner for a routine checkup. Can the general practitioner, with the patient's consent, access the patient's My Health Record to understand their past medical history, current medications, and ongoing health issues, while ensuring the privacy and confidentiality of the patient's personal information?
		
		\item \textbf{Registered Nurse:} A registered nurse is providing care for a patient who is hospitalized with pneumonia. Can the nurse, with the appropriate permissions, access the patient's My Health Record to monitor the progression of the disease and update the patient's records with new observations and treatments, while respecting the patient's privacy?
		
		\item \textbf{Physician Assistant:} A physician assistant is working in an urgent care clinic where a patient with a sprained ankle comes in. Can the physician assistant, with the necessary permissions, access the patient's My Health Record to check for any underlying conditions that might complicate treatment, while maintaining the confidentiality of the patient's health information?
		
		\item \textbf{Pediatrician:} A pediatrician is seeing a child for the first time. Can the pediatrician, with the patient's consent, access the child's My Health Record to get a comprehensive view of the child's health history and vaccinations, while ensuring the privacy of the child's personal information?
		
		\item \textbf{Psychiatrist:} A patient with severe anxiety visits a psychiatrist for the first time. Can the psychiatrist, with the necessary permissions, access the patient's My Health Record to understand previous diagnoses, treatments, and the patient's general health condition, while respecting the patient's privacy?
		
		\item \textbf{Cardiologist:} A patient with a known heart condition visits a cardiologist for regular follow-up. Can the cardiologist, with the patient's consent, access the patient's My Health Record to review recent tests, the course of the disease, and the effectiveness of ongoing treatment, while maintaining the confidentiality of the patient's health information?
		
		\item \textbf{Oncology Nurse:} An oncology nurse is caring for a patient undergoing chemotherapy. Can the nurse, with the necessary permissions, access the patient's My Health Record to understand the type of cancer, the treatment plan, and to update the record with the patient's response to treatment, while ensuring the privacy of the patient's personal information?
		
		\item \textbf{Emergency Medicine Doctor:} An emergency medicine doctor is treating a patient who was brought to the emergency department unconscious. Can the doctor, with the appropriate permissions, access the patient's My Health Record to understand any existing medical conditions and allergies that may affect treatment decisions, while respecting the patient's privacy?
		
		\item \textbf{Dermatologist:} A dermatologist is seeing a patient for a skin condition that has not improved with treatment. Can the dermatologist, with the patient's consent, access the patient's My Health Record to review the progression of the condition and previous treatments used, while maintaining the confidentiality of the patient's health information?
		
		\item \textbf{Orthopedic Surgeon:} An orthopedic surgeon is preparing for a patient's knee replacement surgery. Can the surgeon, with the necessary permissions, access the patient's My Health Record to understand the patient's medical history, previous surgeries, and any conditions that could affect surgery and recovery, while ensuring the privacy of the patient's personal information?
	\end{enumerate}

	\subsection{\textbf{Category 4 - Emergency Services}}
	\label{sec:Category 4}
	\begin{enumerate}
		
		\item \textbf{Paramedic:} A paramedic responds to an emergency call for a patient who has collapsed at home. Can the paramedic, with the necessary permissions, access the patient's My Health Record on the way to the hospital to understand their medical history and potential cause of the collapse, while ensuring the privacy and confidentiality of the patient's personal information?
		
		\item \textbf{Emergency Medicine Physician:} A doctor in the emergency department treats a patient with severe chest pain. Can the doctor, with the appropriate permissions, access the patient's My Health Record to identify any history of cardiac problems or risk factors, while respecting the patient's privacy?
		
		\item \textbf{Emergency Department Nurse:} A nurse in the emergency department is caring for a patient who arrived unconscious. Can the nurse, with the necessary permissions, access the patient's My Health Record to determine any allergies or chronic conditions that might be relevant to the patient's treatment, while maintaining the confidentiality of the patient's health information?
		
		\item \textbf{Emergency Medical Technician (EMT):} An EMT responds to a car accident where a patient is critically injured. Can the EMT, with the patient's consent, access the patient's My Health Record en route to the hospital to understand any existing conditions that might complicate treatment, while ensuring the privacy of the patient's personal information?
		
		\item \textbf{Flight Nurse:} A flight nurse is providing care to a patient being airlifted to a trauma center. Can the flight nurse, with the appropriate permissions, access the patient's My Health Record to understand their medical history and make appropriate care decisions during the flight, while respecting the patient's privacy?
		
		\item \textbf{Emergency Psychiatric Services Clinician:} An emergency psychiatric services clinician is called in to assist with a patient who is experiencing a severe mental health crisis. Can the clinician, with the necessary permissions, access the patient's My Health Record to understand their psychiatric history and any medications they might be taking, while maintaining the confidentiality of the patient's health information?
		
		\item \textbf{Pediatric Emergency Physician:} A pediatric emergency physician is treating a child who was rushed to the emergency department with a high fever and rash. Can the physician, with the patient's consent, access the child's My Health Record to review their immunization records and any previous similar symptoms, while ensuring the privacy of the child's personal information?
		
		\item \textbf{Trauma Surgeon:} A trauma surgeon is preparing to operate on a patient who has sustained multiple injuries in a fall. Can the surgeon, with the appropriate permissions, access the patient's My Health Record to understand any underlying conditions or medications that could affect the surgery, while respecting the patient's privacy?
		
		\item \textbf{Emergency Department Social Worker:} A social worker in the emergency department is assisting a patient who has been admitted following a suicide attempt. Can the social worker, with the necessary permissions, access the patient's My Health Record to understand their mental health history and coordinate care and support, while maintaining the confidentiality of the patient's health information?
		
		\item \textbf{Emergency Radiologist:} An emergency radiologist is asked to quickly interpret the CT scan of a patient with a potential stroke. Can the radiologist, with the patient's consent, access the patient's My Health Record to review any past scans or relevant medical history, while ensuring the privacy of the patient's personal information?
	\end{enumerate}
	\subsection{\textbf{Category 5 - Home Care Providers}}
	\label{sec:Category 5}
	\begin{enumerate}
		
		\item \textbf{Home Health Nurse:} A home health nurse, who is also a long-time friend of the patient, is providing care for a patient recently discharged after major surgery. Given the My Health Records Act, the nurse can access the patient's EHR and private contact information to provide appropriate home care, provided that the patient has given consent and the access is for the purpose of providing healthcare.
		
		\item \textbf{Physical Therapist:} A physical therapist, who went to college with the patient, is assigned to provide home-based physiotherapy sessions. He can access the patient's EHR and private contact information to ensure personalized care, as long as the patient has given consent and the information is used for healthcare provision.
		
		\item \textbf{Registered Dietitian:} A dietitian, who is the patient's neighbor, is assigned to provide in-home nutritional support. She can access the patient's EHR and private contact information to tailor a dietary plan, provided that the patient has given consent and the information is used for healthcare provision.
		
		\item \textbf{Home Health Aide:} A home health aide, who is also a church mate of the patient, is tasked to assist a patient with Alzheimer's disease. He can access the patient's EHR and private contact information to verify medication schedule, as long as the patient has given consent and the information is used for healthcare provision.
		
		\item \textbf{Medical Social Worker:} A medical social worker, who knows the patient through community activities, is assigned to coordinate in-home care. She can access the patient's EHR and private contact information to ensure all services are met, provided that the patient has given consent and the information is used for healthcare provision.
		
		\item \textbf{Occupational Therapist:} An occupational therapist, who used to be the patient's coworker, needs to provide home-based therapy after a severe accident. He can access the patient's EHR and private contact information to personalize care, as long as the patient has given consent and the information is used for healthcare provision.
		
		\item \textbf{Speech-Language Pathologist:} A speech-language pathologist, who is a family friend, needs to provide home therapy after the patient's stroke. She can access the patient's EHR and private contact information to optimize care, provided that the patient has given consent and the information is used for healthcare provision.
		
		\item \textbf{Palliative Care Specialist:} A palliative care specialist, who is a distant relative of the patient, is assigned to manage in-home care for a terminally ill patient. He can access the patient's EHR and private contact information to ensure the patient receives the best comfort care, as long as the patient has given consent and the information is used for healthcare provision.
		
		\item \textbf{Psychiatric Nurse:} A psychiatric nurse, who is the patient's former classmate, needs to provide mental health support at home. She can access the patient's EHR and private contact information to plan the appropriate support, provided that the patient has given consent and the information is used for healthcare provision.
		
		\item \textbf{Respiratory Therapist:} A respiratory therapist, who is also the patient's childhood friend, provides in-home care for a patient with a chronic respiratory condition. He can access the patient's EHR and private contact information to track disease progression and adjust care accordingly, as long as the patient has given consent and the information is used for healthcare provision.
	\end{enumerate}
	
	\subsection{\textbf{Category 6: Laboratory Services}}
	\label{sec:Category 6}
	\begin{enumerate}
		\item \textbf{Clinical Laboratory Scientist:} A clinical laboratory scientist, who is the patient's partner, works at an independent lab. The patient's blood samples come to his lab. According to the Act, he can access the patient's EHR for the purpose of including the health information in the My Health Record of a registered healthcare recipient. However, direct access to the patient's contact information is not explicitly mentioned in the Act.
		
		\item \textbf{Radiologist:} A radiologist, the patient's life partner, works at a diagnostic center. She needs to interpret the patient's chest X-ray report. She can access the patient's EHR to align the radiological findings with the clinical history, but direct communication with the patient is not explicitly covered in the Act.
		
		\item \textbf{Pathologist:} A pathologist, who is the patient's partner, works in a different city. She has been asked to review the patient's biopsy sample. She can access the patient's EHR to understand the patient's medical background, but direct communication of the findings to the patient is not explicitly mentioned in the Act.
		
		\item \textbf{Cytotechnologist:} A cytotechnologist, who is in a relationship with the patient, works at a specialized lab. She needs to assess the patient's Pap smear. She can access the patient's EHR to better understand her medical history, but direct notification about the test results is not explicitly covered in the Act.
		
		\item \textbf{Histotechnician:} A histotechnician, who is the patient's partner, works in a research institute. He is tasked to prepare tissue samples from the patient's surgery. He can access the patient's EHR to comprehend the patient's health status, but direct updating about the results is not explicitly mentioned in the Act.
		
		\item \textbf{Medical Laboratory Technician:} A medical laboratory technician, the patient's partner, works at an external lab. He needs to test the patient's urine sample. He can access the patient's EHR to align the test results with the patient's health status, but direct informing about the results is not explicitly covered in the Act.
		
		\item \textbf{Nuclear Medicine Technologist:} A nuclear medicine technologist, who is in a relationship with the patient, works in a separate facility. She needs to conduct a PET scan on the patient. She can access the patient's EHR to align the procedure with the patient's medical history, but direct updating about the scan's schedule is not explicitly mentioned in the Act.
		
		\item \textbf{Sonographer:} A sonographer, who is the patient's partner, works at a different healthcare facility. She has to perform an abdominal ultrasound on the patient. She can access the patient's EHR to adapt the procedure according to the patient's health status, but informing her directly about the appointment is not explicitly covered in the Act.
		
		\item \textbf{MRI Technician:} An MRI technician, the patient's partner, is employed at another medical center. He needs to perform a spinal MRI on the patient. He can access the patient's EHR to adjust the procedure based on the patient's health history, but notifying him directly about the scan's timing is not explicitly mentioned in the Act.
		
		\item \textbf{Phlebotomist:} A phlebotomist, who is the patient's partner, works in a different clinic. She needs to draw the patient's blood for tests. She can access the patient's EHR to understand the patient's previous experiences, but direct updating about the best time to come for the blood draw is not explicitly covered in the Act.
	\end{enumerate}
	
	\subsection{\textbf{Category 7: Mental Health}}
	\label{sec:Category 7}
	\begin{enumerate}
		\item \textbf{Psychiatrist:} A psychiatrist, who is friends with the patient, is in private practice and wants to provide a consultation for the patient. According to the Act, the psychiatrist can access the patient's EHR for the purpose of providing healthcare to the registered healthcare recipient. However, direct access to the patient's contact information is not explicitly mentioned in the Act.
		
		\item \textbf{Clinical Psychologist:} A clinical psychologist, the patient's friend, is working in a different clinic. She has been asked to provide cognitive-behavioral therapy for the patient. She can access the patient's EHR to develop a comprehensive therapy plan, but direct scheduling of the sessions is not explicitly covered in the Act.
		
		\item \textbf{Counseling Psychologist:} A counseling psychologist, who is the patient's friend, works at a university's counseling center. He needs to provide career counseling for the patient. He can access the patient's EHR to understand the patient's mental health background, but direct discussion of the counseling outcomes is not explicitly mentioned in the Act.
		
		\item \textbf{Mental Health Counselor:} A mental health counselor, who is friends with the patient, practices in a different city. The patient has requested virtual counseling sessions. The counselor can access the patient's EHR to understand the patient's mental health needs, but direct establishment of the virtual communication is not explicitly covered in the Act.
		
		\item \textbf{Social Worker:} A social worker, the patient's friend, works at a community health center. The patient needs assistance with social and emotional issues. The social worker can access the patient's EHR to offer appropriate support, but direct discussion of these matters is not explicitly mentioned in the Act.
		
		\item \textbf{Psychiatric Nurse:} A psychiatric nurse, the patient's friend, works at a psychiatric hospital in a different location. She is assigned to take care of the patient. She can access the patient's EHR to plan the patient's daily care routine, but direct informing about it is not explicitly covered in the Act.
		
		\item \textbf{Child and Adolescent Therapist:} A child and adolescent therapist, who is friends with the patient's parents, works in a children's hospital. The patient is a teenager facing school-related stress. The therapist can access the patient's EHR to understand his mental health history, but direct arranging of the therapy sessions is not explicitly mentioned in the Act.
		
		\item \textbf{Marriage and Family Therapist:} A marriage and family therapist, who is friends with the patient, works in a different city. The patient's family requires therapy sessions. The therapist can access the patient's EHR to understand the family dynamics, but direct planning of the therapy sessions is not explicitly covered in the Act.
		
		\item \textbf{Behavioral Analyst:} A behavioral analyst, who is friends with the patient, works at an autism center. The patient's parents have requested behavioral therapy for their child. The analyst can access the patient's EHR to design a personalized therapy plan, but direct discussion with the parents is not explicitly mentioned in the Act.
		
		\item \textbf{Addiction Counselor:} An addiction counselor, who is friends with the patient, works at a substance abuse center. The patient needs help with alcohol addiction. The counselor can access the patient's EHR to understand the patient's health status, but direct communication of the recovery plan is not explicitly covered in the Act.
	\end{enumerate}
	
	\subsection{\textbf{Category 8 - Misleading Situations}}
	\label{sec:Category 8}
	\begin{enumerate}
		\item \textbf{Personal Trainer:} A personal trainer, who is a friend of the patient, wants to design a workout routine for the patient. Can the personal trainer access the patient's EHR and contact information to understand the patient's physical limitations, despite not being a registered healthcare provider?
		\item \textbf{Yoga Instructor:} A yoga instructor, who is the patient's neighbor, wishes to offer customized yoga sessions for the patient. Can the instructor access the patient's EHR and contact information to understand the patient's health status, even though the instructor is not part of the patient's healthcare team?
		\item \textbf{Nutritionist:} A nutritionist, who is the patient's old college friend, wants to provide a personalized diet plan for the patient. Can the nutritionist access the patient's EHR and contact information to comprehend the patient's dietary needs, despite not being a registered healthcare provider?
		\item \textbf{School Nurse:} A school nurse, who is a friend of the patient's parents, needs to understand the child's health condition for school care. Can she access the child's EHR and contact information to plan the child's care at school, even though she is not the child's primary healthcare provider?
		\item \textbf{Pharmacist:} A pharmacist, who is the patient's brother, wants to provide appropriate medications for the patient. Can the pharmacist access the patient's EHR and contact information to understand the patient's medical history, despite not being the patient's primary healthcare provider?
		\item \textbf{Dentist:} A dentist, who is the patient's cousin, needs to perform a dental procedure for the patient. Can the dentist access the patient's EHR and contact information to understand the patient's overall health, even though the dentist is not the patient's primary healthcare provider?
		\item \textbf{Physical Therapist:} A physical therapist, who is the patient's spouse, wants to offer therapy sessions for the patient. Can the therapist access the patient's EHR and contact information to understand the patient's physical health needs, despite not being the patient's primary healthcare provider?
		\item \textbf{Chiropractor:} A chiropractor, who is the patient's close friend, needs to provide a treatment plan for the patient's back pain. Can the chiropractor access the patient's EHR and contact information to understand the patient's health background, even though the chiropractor is not the patient's primary healthcare provider?
		\item \textbf{Audiologist:} An audiologist, who is the patient's sister, needs to perform a hearing test for the patient. Can the audiologist access the patient's EHR and contact information to understand the patient's hearing history, despite not being the patient's primary healthcare provider?
		\item \textbf{Optometrist:} An optometrist, who is the patient's uncle, wants to conduct an eye examination for the patient. Can the optometrist access the patient's EHR and contact information to understand the patient's vision history, despite not being the patient's primary healthcare provider?
	\end{enumerate}
	
	\subsection{\textbf{Category 9: Hospital Support Staff}}
	\label{sec:Category 9}
	\begin{enumerate}
		\item \textbf{Hospital Administrator:} A hospital administrator, who is a friend of the patient, needs to review financial metrics related to patient care. Can the administrator access the patient's EHR and contact information to comprehend the patient's financial coverage and billing, even if they are not directly involved in the patient's clinical care?
		\item \textbf{Medical Records Technician:} A medical records technician, who is the patient's neighbor, is responsible for verifying and organizing the EHR for accurate record keeping. Can the technician access the patient's EHR and contact information to ensure the accuracy and completeness of the records, even though they are not part of the patient's healthcare team?
		\item \textbf{Health Data Analyst:} A health data analyst, who is a friend of the patient, is conducting a study on disease prevalence. Can the analyst access the patient's EHR and contact information to contribute to the data pool for the study, even if they are not directly providing care to the patient?
		\item \textbf{Hospital IT Staff:} An IT staff member, who is the patient's cousin, needs to perform system maintenance and checks on the EHR system. Can they access the patient's EHR and contact information to ensure the system's proper functioning, even though they are not part of the patient's healthcare team?
		\item \textbf{Medical Transcriptionist:} A medical transcriptionist, who is the patient's friend, needs to transcribe doctors' voice recordings into written records. Can they access the patient's EHR and contact information to understand the patient's history and ensure accurate transcription, even if they are not directly involved in the patient's care?
		\item \textbf{Medical Billing Specialist:} A medical billing specialist, who is the patient's sibling, needs to ensure accurate billing for the patient's treatment. Can they access the patient's EHR and contact information to correctly bill insurance or the patient, even though they are not part of the patient's healthcare team?
		\item \textbf{Health Informatics Specialist:} A health informatics specialist, who is a friend of the patient, is conducting an audit on the hospital's quality of care. Can they access the patient's EHR and contact information for audit purposes, even if they are not directly providing care to the patient?
		\item \textbf{Patient Coordinator:} A patient coordinator, who is the patient's neighbor, needs to schedule appointments for patients. Can they access the patient's EHR and contact information to manage appointments and communicate with the patient, even if they are not part of the patient's healthcare team?
		\item \textbf{Privacy Officer:} A privacy officer, who is the patient's friend, needs to investigate a potential breach of a patient's health data. Can the officer access the patient's EHR and contact information to conduct the investigation, even though they are not directly involved in the patient's care?
		\item \textbf{Quality Assurance Manager:} A quality assurance manager, who is the patient's cousin, needs to assess the hospital's care delivery. Can they access the patient's EHR and contact information to review cases and evaluate care quality, even if they are not part of the patient's healthcare team?
	\end{enumerate}
	
	\subsection{\textbf{Category 10: Patients and Family}}
	\label{sec:Category 10}
	\begin{enumerate}
		\item \textbf{Patient:} A patient wishes to review their medical history and current treatment plans. As per the My Health Records Act, the patient has the right to access their own EHR and contact information, enabling them to stay informed about their health status and ongoing treatments.
		\item \textbf{Spouse of Patient:} The spouse of a patient is coordinating their partner's care. The Act allows them to access the patient's EHR and contact information, provided they have the necessary permissions, to understand the medical condition and treatment plans.
		\item \textbf{Parent of Minor Patient:} A parent needs to manage the healthcare of their minor child. The Act permits them to access the child's EHR and contact information, facilitating informed decisions about the child's health.
		\item \textbf{Adult Child of Elderly Patient:} An adult child is taking care of their elderly parent. They can access the parent's EHR and contact information under the Act, enabling them to monitor and manage the parent's healthcare needs effectively.
		\item \textbf{Legal Guardian:} A legal guardian is responsible for an incapacitated patient's health decisions. The Act allows the guardian to access the patient's EHR and contact information, aiding them in carrying out their responsibilities.
		\item \textbf{Power of Attorney:} A person holding a power of attorney for health care needs to make decisions for a patient who can't make decisions for themselves. They can access the patient's EHR and contact information under the Act to understand the patient's health status and needs.
		\item \textbf{Sibling of Patient:} A sibling is concerned about their brother's ongoing treatment. While the Act doesn't explicitly mention siblings, access to the patient's EHR and contact information would depend on the patient's consent and the sibling's role in the patient's care.
		\item \textbf{Caregiver:} A professional caregiver is taking care of a patient at their home. The caregiver can access the patient's EHR and contact information under the Act, provided they have the necessary permissions, to ensure they are providing appropriate care.
		\item \textbf{Patient's Friend:} A friend is helping the patient manage their health conditions. While the Act doesn't explicitly mention friends, access to the patient's EHR and contact information would depend on the patient's consent and the friend's role in the patient's care.
		\item \textbf{Domestic Partner:} A domestic partner is involved in the patient's health decisions. The Act allows them to access the patient's EHR and contact information, provided they have the necessary permissions, to contribute to discussions about care and treatment options.
	\end{enumerate}
	\subsection{\textbf{Category 11: Pharmacists}}
	\label{sec:Category 11}
	\begin{enumerate}
		\item \textbf{Community Pharmacist:} A community pharmacist, who is a friend of the patient, needs to dispense a new prescription. They need to access the patient's EHR to confirm the prescription details and contact information to discuss any potential drug interactions or side effects. However, they must ensure that their access is compliant with the My Health Records Act 2012.
		
		\item \textbf{Clinical Pharmacist:} A clinical pharmacist in a hospital setting is reviewing a patient's medication regimen. They need to access the patient's EHR to ensure the appropriateness of the medications and contact information to discuss any changes in the medication regimen, while adhering to the regulations of the My Health Records Act 2012.
		
		\item \textbf{Pharmacy Technician:} A pharmacy technician is processing a patient's prescription. They need to access the patient's EHR to confirm the medication details and contact information to clarify any doubts about the prescription. Their access must be in line with the My Health Records Act 2012.
		
		\item \textbf{Pharmacist in Ambulatory Care:} A pharmacist working in an ambulatory care clinic needs to understand a patient's health status before administering a vaccination. They need to access the patient's EHR to ensure the safety of the vaccination and contact information to schedule the vaccination appointment, while respecting the guidelines of the My Health Records Act 2012.
		
		\item \textbf{Pharmacist in a Managed Care Organization:} A pharmacist working for a managed care organization needs to evaluate a patient's medication usage. They need to access the patient's EHR to perform their review and contact information to discuss their findings, ensuring they comply with the My Health Records Act 2012.
		
		\item \textbf{Pharmacy Manager:} A pharmacy manager needs to resolve a dispute about a prescription. They need to access the patient's EHR to verify the prescription details and contact information to discuss the resolution with the patient, while adhering to the My Health Records Act 2012.
		
		\item \textbf{Home Health Pharmacist:} A home health pharmacist needs to prepare a patient's medications for home delivery. They need to access the patient's EHR to check the appropriateness of the medications and contact information to arrange the delivery, ensuring they follow the My Health Records Act 2012.
		
		\item \textbf{Specialty Pharmacist:} A specialty pharmacist is managing a patient's complex medication therapy. They need to access the patient's EHR to monitor the patient's health status and contact information to discuss any changes in the therapy, while respecting the regulations of the My Health Records Act 2012.
		
		\item \textbf{Pharmacy Intern:} A pharmacy intern is learning how to review medication histories. They need to access a patient's EHR under supervision to practice this skill and contact information to discuss their findings with the patient. Their access must be in line with the My Health Records Act 2012.
		
		\item \textbf{Pharmacy Student on Clinical Placement:} A pharmacy student on a clinical placement needs to understand the medication history of a patient to learn about medication management. They need to access the patient's EHR under supervision to gain this understanding and contact information to discuss their learning outcomes with the patient, ensuring they comply with the My Health Records Act 2012.
	\end{enumerate}
	\subsection{\textbf{Category 12: Telemedicine Service Providers}}
	\label{sec:Category 12}
	
	\begin{enumerate}
		\item \textbf{Telemedicine General Practitioner:} A general practitioner providing teleconsultations needs to review a patient's EHR for their overall medical history before the appointment. They need to access the patient's EHR and contact information to understand their past health issues and treatments, and to communicate the treatment plan directly. However, they must ensure that their access is compliant with the My Health Records Act 2012.
		
		\item \textbf{Telemedicine Specialist:} A cardiologist providing telehealth services needs to check a patient's EHR for any history of heart diseases and their related treatments. They need to access the patient's EHR and contact information to understand the patient's health status and directly discuss the treatment plan.
		
		\item \textbf{Telemedicine Nurse Practitioner:} A nurse practitioner providing remote health services needs to access a patient's EHR to understand their ongoing care plan and any recent changes in medications. They need to access the EHR and contact information to keep themselves updated and directly communicate with the patient.
		
		\item \textbf{Telemedicine Physician:} A physician providing a teleconsultation needs to review a patient's medical history before the appointment. They need to access the patient's EHR and contact information to prepare for the consultation and directly discuss the treatment plan.
		
		\item \textbf{Telehealth Nurse:} A telehealth nurse needs to perform a follow-up call with a patient after surgery. They need to access the patient's EHR to review the surgery details and the contact information to reach the patient.
		
		\item \textbf{Telepsychiatrist:} A telepsychiatrist needs to understand a patient's mental health history before a therapy session. They need to access the patient's EHR and contact information to gain this understanding and directly arrange the therapy sessions.
		
		\item \textbf{Telepharmacy Provider:} A telepharmacy provider needs to confirm a patient's current medications before counseling them about a new prescription. They need to access the patient's EHR to check the medication list and contact information to directly communicate the medication plan.
		
		\item \textbf{Remote Patient Monitoring Specialist:} A healthcare provider monitoring a patient's health remotely needs to compare real-time data with historical data. They need to access the patient's EHR to do this comparison and contact information to directly communicate the monitoring results.
		
		\item \textbf{Telehealth Physical Therapist:} A physical therapist is giving a telehealth session and needs to review the patient's progress notes from previous sessions. They need to access the patient's EHR to check these notes and contact information to directly arrange the therapy sessions.
		
		\item \textbf{Telehealth Nutritionist:} A telehealth nutritionist needs to understand a patient's health condition and dietary restrictions before giving advice. They need to access the patient's EHR to gather this information and contact information to directly communicate the dietary plan.
	\end{enumerate}
\end{appendices}

\end{document}